\documentclass{aa}  

\usepackage{graphicx}
\usepackage{hyperref}
\hypersetup{colorlinks=True, citecolor={blue}, urlcolor={blue},linkcolor={black}}
\usepackage[capitalise]{cleveref}
\crefname{section}{Sect.}{Sects.}
\usepackage{txfonts}
\usepackage{stackengine}
\def\cross{%
  \stackon[1ex]{\rule{0.4pt}{1.5ex}}{\rule{.75ex}{0.4pt}}}
\usepackage{booktabs}
\usepackage{threeparttable}

\newcommand{\threesigma}{$3\sigma_{\text{rms}}$\xspace}

\newcommand{\CIZA}{CIZA J2242.8+5301\xspace}
\newcommand{\ainj}{$\alpha_\mathrm{inj}$\xspace}
\newcommand{\aint}{$\alpha_\mathrm{int}$\xspace}

\newcommand{\reviewfirst}[1]{\textcolor{black}{#1}} 

\newcommand{\reviewsecond}[1]{\textcolor{black}{#1}} 

\makeatletter
\let\linenumbers\relax
\let\nolinenumbers\relax
\makeatother

\begin{document} 

   \title{A view of the CIZA J2242.8+5301 galaxy cluster at very low radio frequencies}

   \subtitle{}

   \author{G. Lusetti
          \inst{1}
          \and
          M. Br\"uggen\inst{1}
          \and
          H. W. Edler\inst{1}
          \and
          F. de Gasperin\inst{2,1}
          \and
          M. Hoeft\inst{3}
          \and
          G. Di Gennaro\inst{2}
          \and
          D. Hoang\inst{3}
          \and
          T. Pasini\inst{2}
          \and
          R. van Weeren\inst{4}
          \and
          V. Cuciti\inst{2}
          \and
          H. Rottgering\inst{4}
          \and
          G. Brunetti\inst{2}
          }

   \institute{
        Hamburger Sternwarte, University of Hamburg, Gojenbergsweg 112, 21029 Hamburg, Germany \\
            \email{giulia.lusetti@hs.uni-hamburg.de}
    \and 
        INAF - Istituto di Radioastronomia di Bologna, Via Gobetti 101, 40129 Bologna, Italy
    \and 
        Thüringer Landessternwarte, Sternwarte 5, 07778 Tautenburg, Germany
    \and 
        Leiden Observatory, Leiden University, PO Box 9513, 2300 RA Leiden, The Netherlands 
             }

   \date{2024}

  \abstract
   {The galaxy cluster CIZA J2242.8+5301 is a well-studied merging galaxy cluster that hosts prominent double radio relics including the famous sausage relic, as well as other diffuse radio sources.   
   Observations at frequencies below 100 MHz are essential for investigating the physics of radio relics as they provide unique access to the low-energy population of cosmic-ray electrons. 
   }
   {We aim to study the morphology, spectral characteristics, and physical processes that produce relics.}
   {We present the first observations of the Sausage cluster at 45 MHz, the lowest radio frequency at which this cluster has been studied to date, using the Low Band Antenna (LBA) of the LOFAR radio interferometer. We made use of ten hours of LOFAR LBA observations, from which we achieved a thermal-noise limited radio image with a noise level of $ \rm 1.5 \, mJy/beam $ at a resolution of $ 15\arcsec $. 
   These data were combined with existing multi-frequency measurements at higher frequencies: 
   LOFAR High Band Antenna (HBA: 145 MHz); Giant Metrewave Radio Telescope (GMRT: 325, 610 MHz); Westerbork Synthesis Radio Telescope (WSRT: 1.2, 1.4 GHz); and Karl G. Jansky Very Large Array (VLA: 1.5, 3 GHz).
   This broad frequency coverage allowed us to derive integrated spectral indices, spectral index and curvature maps, and Mach number distributions across the relics.
}
   {We derived Mach numbers from the local injection index measure using low-frequency data with $\mathcal{M}_N = 2.9 \pm 0.5$ for the northern relic and $\mathcal{M}_S = 2.9 \pm 0.8$ for the southern relic. 
    LOFAR LBA observations reveal a remarkably symmetric surface brightness profile across the eastern part of the northern relic, with wings extending on either side of the peak. This discovery is contrary to the expectation of particle acceleration at a single, sharp shock and the subsequent downstream advection of accelerated electrons. 
    We modelled the surface brightness profile, including the effects of projection, magnetic field variation, and shock deformation.
 }
   {}


   \maketitle

\section{Introduction}

During the formation of the large-scale structure, galaxy clusters emerge through hierarchical merging and accretion processes. These events release vast amounts of energy into the intra-cluster medium (ICM), generating shocks and turbulence that can accelerate particles to relativistic energies \citep{Brunetti&Jones2014}. 
When relativistic electrons interact with the cluster’s $\mu G$ magnetic field, they emit synchrotron radiation, making these processes observable at radio wavelengths. Evidence of these processes is widely observed in the form of Mpc-scale diffuse radio sources, such as radio halos and relics \citep{vanWeeren2019}.
Radio halos are diffuse, centrally located, steep spectrum sources that roughly follow the ICM distribution. They are believed to be powered by turbulence induced by cluster mergers, which re-accelerates a pre-existing population of relativistic electrons \citep{Brunetti&Jones2014}.
Radio relics are elongated synchrotron sources located at the outskirts of galaxy clusters \citep[see][for a recent review]{Wittor2023Univ}. 
They generally exhibit a bright outer rim in radio emission, which fades towards the cluster centre. This gradient in radio intensity is often accompanied by a steepening spectral index: from a relatively flat value at the edge of the relic, which indicates freshly accelerated electrons, to a steeper value closer to the cluster core, where the electrons lose energy through synchrotron and inverse-Compton (IC) radiative processes.
When observed edge-on, these structures often display an arc-like shape, resembling the shocks believed to produce them. However, high-sensitivity radio observations have revealed that radio relics can adopt a variety of morphologies, which complicates our understanding of their origins and evolution. An increasing number of studies show relics with filamentary structures \citep[e.g.][]{deGasperin2022}, inverted morphology \citep[e.g.][]{Riseley2022}, or almost completely face-on \citep[e.g.][]{Rajpurohit2022_A2256}.
This diversity in shape is also supported by the latest simulations \citep{Lee2024}, which show that relics are not smooth structures but are intrinsically made of filaments and display a variety of morphologies driven by the complex dynamics of the ICM.

The prevailing theory is that radio relics form through the shock acceleration of cosmic-ray electrons via the diffusive shock acceleration (DSA) process \citep{BlandfordOstriker1978, Drury1983, Blandford1987,Ensslin1998, HoeftBruggen2007}. These shocks arise during cluster mergers, as large substructures collide and pass through one another, compressing and heating the ICM.
The standard DSA theory involves the thermal electrons being directly accelerated (sometimes referred to as fresh injection).
However, a major challenge is that the observed low Mach numbers ($\rm \mathcal{M}\lesssim3$) would imply unrealistically large particle injection efficiencies, which DSA from the thermal pool alone hardly explains \citep[e.g.][]{Botteon2020_shock_acceleration}. To solve this tension, models have invoked the presence of a pre-existing population of mildly relativistic electrons \citep[e.g.][]{Markevitch2005_A520,Kang_and_Ryu2011, Pinzke2013} either pre-accelerated by previous shock passages \citep{Inchingolo2022} or coming from fossil plasma from radio galaxies, often observed in the proximity or radio relics \citep[e.g.][]{vanWeeren2017NatAs}.
However, it is not straightforward that the injection of fossil electrons by one or more radio galaxies would automatically produce a uniform population of electrons, able to produce the highly coherent radio emission observed in some giant relics \citep[e.g.][]{vanWeeren2010, vanWeeren2012_toothbrush}.
Another observational challenge has been the Mach number discrepancy. The shock Mach number can be inferred from both X-ray (via shock jump conditions) and radio (via the spectral slope of synchrotron emission) observations. In principle, these two independent methods should be consistent, as they measure the same shock. However, they rarely agree, with $\rm \mathcal{M}_{radio}>\mathcal{M}_{X-ray}$.
Recent simulations \citep{Wittor2021, Whittingham2024} resolved this inconsistency, showing that the discrepancy naturally arises from the differences in how the two methods sample the underlying Mach number distribution \citep{Hoeft2011}.
Radio observations are more sensitive to localized regions of high Mach numbers, which only characterize a small fraction of the shock’s surface, while X-ray observations measure an average of the Mach number distribution across the entire shock front, which leads to lower inferred values.\\

Studying individual relics in detail is crucial for testing and refining current models. One of the most studied systems in this regard is \CIZA, which has become an ideal benchmark system due to its well-defined relic structures and wealth of multi-wavelength data.  
CIZA J2242.8+5301 is a dissociative major merging galaxy cluster located at redshift $ z = 0.1921$. It was first identified in X-rays by \cite{Kocevski2007_CIZA} using \textit{ROSAT}, but became widely known for hosting one of the most striking examples of a double radio relic. In particular, the distinctive morphology of the northern relic led to its designation as the 'Sausage cluster' \citep{vanWeeren2010}. Due to its high brightness and morphological simplicity, the Sausage galaxy cluster quickly became a preferred target for the study of shock acceleration in the ICM.

Over the last decade, it has been extensively studied in the radio band across a wide frequency range, from 150 MHz to 30 GHz \citep{vanWeeren2010, vanWeeren2011GMRT, 2013Stroe, Stroe2014_16GHz, Stroe2016, Hoang2017, Loi2017, DiGennaro2018, Loi2020, DiGennaro2021, Raja2024}.
However, deeper observations have revealed that the radio emission in \CIZA is more complex than initially thought. While the cluster hosts a seemingly simple double-relic system, at approximately 1.5 Mpc north and south from the cluster centre, additional relic-like sources are detected in the downstream region of the main relics, as well as on the adjacent eastern and western sides of the radio shocks. This observationally proves that the merger environment is more intricate than a straightforward binary collision scenario, with multiple shock interactions and possible re-accelerated plasma components shaping the observed radio structures.  
In order to better characterize the shape of the relic spectrum, observations at frequencies up to 30 GHz have been performed. 
These observations aimed to investigate whether the integrated spectrum follows a single power law, as predicted by the DSA model.
Initially, \cite{Stroe2016} found a significant steepening of the spectrum from $\sim-1.0$ to $\sim-1.6$, above 2 GHz using interferometric data, and suggested that this could be due to contributions from the Sunyaev-Zel’dovich (SZ) effect or the presence of a non-uniform magnetic field. 
However, subsequent single-dish observations with the Effelsberg Telescope \citep{Kierdorf2017}, as well as combined single-dish and interferometric measurements from the Sardinia Radio Telescope \citep{Loi2017}, found no evidence of spectral steepening at high frequencies.
The cluster contains a low surface brightness radio halo (with $\rm \bar{\alpha}^{145MHZ}_{2.3GHz}=-1.01\pm0.10$; \citealt{Hoang2017}), previously also reported by \citet{vanWeeren2010}, \citet{Stroe2013Discovery}, and \citet{DiGennaro2018}. The radio halo characterization proved to be difficult, due to its connection to the relics emission and the substantial contamination from tailed radio galaxies.\\

Many efforts have been devoted to studying \CIZA in the X-ray band as well \citep{Ogrean2013XMM, Akamatsu2013, Ogrean2014Chandra, Akamatsu2015}, which is crucial for confirming the presence of shocks and investigating the dynamics of the merger scenario.
Analyses by \cite{Ogrean2013XMM} with \textit{XMM-Newton} and by \cite{Akamatsu2013} with \textit{Suzaku} revealed an elongated X-ray morphology aligned with the merger axis indicated by the radio relics.
However, detecting an unambiguous cluster merger shock (exhibiting both a sharp gas density edge and a temperature jump) in connection with the relics has proven challenging. This is partly due to the low resolution of Suzaku and XMM-Newton, and partly because the relics reside in low signal-to-noise regions of the instrumental field of view.
\cite{Akamatsu2013} observed a temperature jump at the northern relic corresponding to a Mach number of 3.15, but no surface brightness jump in the X-ray profiles.
These authors showed that this is due to the limited spatial resolution of Suzaku, which is larger than the length of the shock, causing the discontinuity to be diluted and thus undetectable.
Also \cite{Ogrean2013XMM} found no evidence of shock compression near the northern relic, while they identified a surface brightness jump by a factor of $\rm\sim 2-3$, corresponding to a weak shock with a Mach number of 1.2-1.3, near the southern one (approximately 1 Mpc from the centre).
However, due to the size of the XMM-Newton field-of-view (FoV), even in this case it was not possible to get data immediately beyond the southern relic to better constrain the gas properties at the exact relic position.
In a consequential study, \cite{Ogrean2014Chandra} found multiple density discontinuities spread throughout the cluster volume but not a clear, strong shock signature at the northern relic location. While they confirmed the detection of a temperature jump (consistent with a Mach number of 2.54), they only found a hint ($\rm <2\sigma$ detection) of a density jump at the northern relic location using Chandra data. Finally, \cite{Akamatsu2015} managed to detect two clear temperature drops at the location of the northern and southern relic, finding Mach values of 2.7 and 1.7, respectively.
\\

Optical studies confirmed that the dynamics of the merger is dominated by two sub-clusters with comparable masses, leading to a near 1:1 mass ratio. In particular, spectroscopic velocity-dispersion analysis \citep{Dawson2015} derived $\rm M_{1}= 16.1_{-3.3}^{+4.6}\times 10^{14}M_{\odot}$ and $\rm M_{2} = 13.0_{-2.5}^{+4.0}\times 10^{14}M_{\odot}$ for the northern and southern sub-clusters respectively, while weak-lensing inferred masses \citep{Jee2015} were $\rm M_{200} = 11.0^{+3.7}_{-3.2} \times 10^{14}M_{\odot}$ and $9.8^{+3.8}_{-2.5} \times 10^{14} M_{\odot}$ for the northern and southern halos, respectively.
Furthermore, the low, relative line-of-sight velocity of $\rm 69\pm190 \,km/s$ \citep{Dawson2015} supports the scenario of a merger occurring close to the plane of the sky.
The excellent agreement between the merger axis inferred from radio relics, ICM, and bi-modal galaxy distribution strongly suggests that the merger is occurring close to the plane of the sky.\\

Being one of the brightest and most widely studied galaxy clusters, the Sausage cluster offers a unique opportunity to study the properties of radio relics at very low radio frequencies ($\lesssim 100$ MHz) and to prove the technical capability of the LOw Frequency ARray \citep[LOFAR;][]{vanHaarlem2013} telescope and its related data calibration.
Moreover, the small projection effect, clear shock feature, and the large amount of observational data available make this object a perfect reference target in simulation and for testing analytical models \citep{Donnert2016, Kierdorf2017, Kang2016}.\\

In this paper, we present  45 MHz observations of CIZA J2242.8+530, conducted using the LOFAR Low Band Antennas (LBA), for the first time. 
We complement our study by including higher frequency data from \citet{Hoang2017}, \citet{Stroe2013Discovery}, \citet{vanWeeren2010}, and \citet{DiGennaro2018}.
The paper is organized as follows. We describe the radio observations and the data reduction in \cref{sec:data_reduction}. In \cref{sec:results}, we present  our results for the Sausage galaxy cluster, including its morphology, spectral index, and curvature analysis. The discussion appears in \cref{sec:discussion}, including a thorough study of the shock Mach number and surface brightness profile modelling of the northern relic. Finally, we summarize our conclusions in \cref{sec:conclusion}. 

Throughout this paper, we adopt a $\Lambda$CDM cosmology with $H_0 = 70\,\mathrm{km\,s^{-1}\,Mpc^{-1}}$, $\Omega_\mathrm{m} = 0.3$, and $\Omega_\Lambda = 0.7$. At the redshift of the Sausage cluster ($z \approx 0.19$), we have $1^{\arcsec} = 3.169$ kpc.  
The spectral index is defined as $S_\nu \propto \nu^{\alpha}$, where $S_\nu$ is the flux density.
The uncertainty, $\Delta_S$ , associated with a flux density measurement, $S,$ is estimated as

\begin{equation}
    \Delta_S = \sqrt{  (\sigma_{\rm c}\cdot S)^2 + N_{\rm beam}\cdot \sigma_{\rm rms}^2} ,
    \label{eq:flux_error}
\end{equation}
where $N_{\rm beam}=N_{\rm pixel}/A_{\rm beam}$ is the number of independent beams in the source area, and $\sigma_{\rm c}$ indicates the systematic calibration error on the flux density, with a typical value of 10\% for LOFAR LBA \citep{deGasperin2021}.

\section{Observation and data reduction}
\label{sec:data_reduction}

In this section, we describe the data reduction for the LOFAR LBA data. In \cref{tab:radio_info}, we summarize the details of the observations.
\reviewfirst{Although this data are part of the project LC18\_13 (PI: F. de Gasperin), which also included the LOFAR International Stations, for this work only the Dutch array (Core and Remote Stations) was used.}\\

\begin{table}[h]
\centering
\caption{Observation details.}
\begin{tabular}{ll}
\hline
Target          &    CIZA J2242.8+5301\\
Calibrator      &    3C380 \\
Observing time  &  5th July 2022, 5h \\
                &  7th July 2022, 5h \\
Frequency range &  20-68MHz \\
Time resolution*        &  4 s \\
Frequency resolution*   &  8 ch/SB \\
Antenna-set     & LBA\_SPARSE \\
Project**         & LC18\_13 \\
Obs. ID         & 2005118 \\
                & 2005123 \\
                
\hline
\end{tabular}
\begin{tablenotes}
    \item[*] *: after initial averaging. Sub-band bandwidth: 0.195 MHz.
    \reviewfirst{\item[*] **: These observations were obtained solely with the Dutch array; no international stations were used. }
\end{tablenotes}
\label{tab:radio_info}
\end{table}
   
\subsection{Demixing}

\begin{figure*}
    \includegraphics[width=\textwidth]{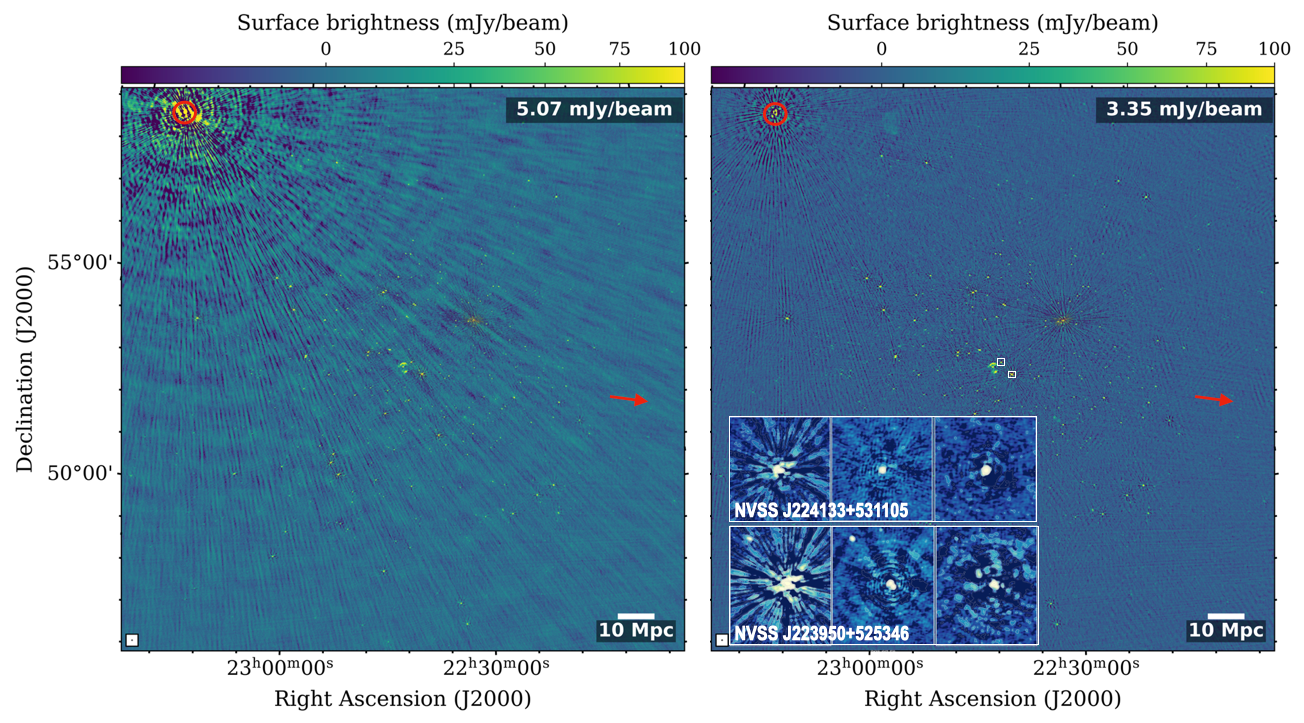}
    \caption{Results of the demix without considering (left) and considering (right) the target field sky model. 
    The red circle indicates the location of Cassiopeia A while the arrow points towards the location of Cygnus A outside the FoV.
    The Sausage cluster resides at the centre of the image, the two bright radio galaxies NVSS J224133+531105 and NVSS J223950+525346 are highlighted with white squares. The insets show zoom-in images on these two radio galaxies after DI, DD, and extraction calibration steps, from left to right respectively.}
    \label{fig:demix-results}
\end{figure*}

The target lies about 8 and 30 deg away from CasA ($\rm\sim 27\,000\,Jy$ at 50 MHz) and CygA ($\rm\sim 22\,000\,Jy$ at 50 MHz). Being among the brightest radio sources on the sky, it is necessary to carefully remove these sources from the data to avoid strong contamination by their emission spilling into LOFAR sidelobes. For this, we employ the demixing algorithm \citep{VanderTol2009PhDT}, a technique specifically designed to remove the contribution of ultra-bright off-axis sources. This allows us to calibrate and subtract the bright sources from the visibilities. 
The model used for CasA and CygA is obtain from $\sim10\arcsec$ resolution images \citep{deGasperin2020_ATeam} and contains over $1600$ and $300$ components, respectively.
To accurately subtract the sources, we found that it is critically important to also take into account the sky model of the sources in the target field during demixing. This is currently not routinely done in pre-processing of LOFAR observations at the level of the LOFAR Observatory.
Our model of the target field was constructed from the Global Sky Model which is built from radio surveys  such as TGSS (TIFR GMRT Sky Survey, \cite{Intema2017}), NVSS (NRAO VLA Sky Survey, \cite{Condon1998}), WENSS (Westerbork Northern Sky Survey, \cite{Condon1998}) and VLSSr (Very Large Array Low-frequency Sky Survey Redux, \cite{Lane2014}).
We show the impact of the demixing process with and without a target field sky model in \cref{fig:demix-results}.

\subsection{Cross- and self-calibration}

The LBA observations were performed in observation mode LBA\_SPARSE and using the multi-beam capability of LOFAR, which allows to continuously point one beam towards the calibrator source 3C380.
For calibration of the LBA data, we employ the standard routines implemented in the Library for Low-Frequencies (\texttt{LiLF} \footnote{\url{https://github.com/revoltek/LiLF}}).
The initial step involves computing the solutions for the calibrator as described in \citet{deGasperin2019}. 
The calibration solutions were then applied to the target field data together with an analytic model of the dipole beam effect.

Next, the direction-independent (DI) self calibration of the target field was performed, as presented in \cite{deGasperin2020}. This step aims at finding the solutions for the average ionospheric effects. 
From an initial sky model derived from the LOFAR Global Sky Model, solutions for the total electron content (TEC), Faraday rotation and second-order beam effects are derived and applied.

The last step consists of direction-dependent (DD) calibration, following \cite{deGasperin2020, Edler2022}. This step addressed the remaining direction-dependent errors caused by the ionosphere. 
Firstly, the Python blob detector and source finder \citep[\texttt{PyBDSF};][]{Mohan2015} was used to select bright and compact sources, which are then used as direction-dependent calibrators. Then, all sources of the source model found after direction-independent calibration were subtracted from the $uv$-data to create an empty data set. One at a time starting with the brightest source, the calibrators were re-added to the empty data, the data was phase-shifted to the calibrator direction and averaged in time and frequency, corrected for the primary beam in the new phase centre and self-calibrated in several cycles. If the rms noise level of the image did not improve by more than 1\% during a self-calibration cycle, the corresponding calibrator direction was considered converged, the calibrator was re-subtracted from the data set using the improved source model and solutions and the calibration for the next calibrator started. 
We performed a single DD cycle, that allowed us to obtain a high-fidelity image with well-recovered compact and extended emission.

\subsection{Extraction}

To further improve the quality of the image at the target position, in particular to reduce the artifacts in the vicinity of the strong sources, we employed the extraction technique. This was originally developed by \cite{vanWeeren2021extraction} for LOFAR HBA data. Here we use and describe the LOFAR-LBA adapted approach described by \cite{Pasini2022}.

\begin{figure*}[h]
    \includegraphics[width=0.9\textwidth]{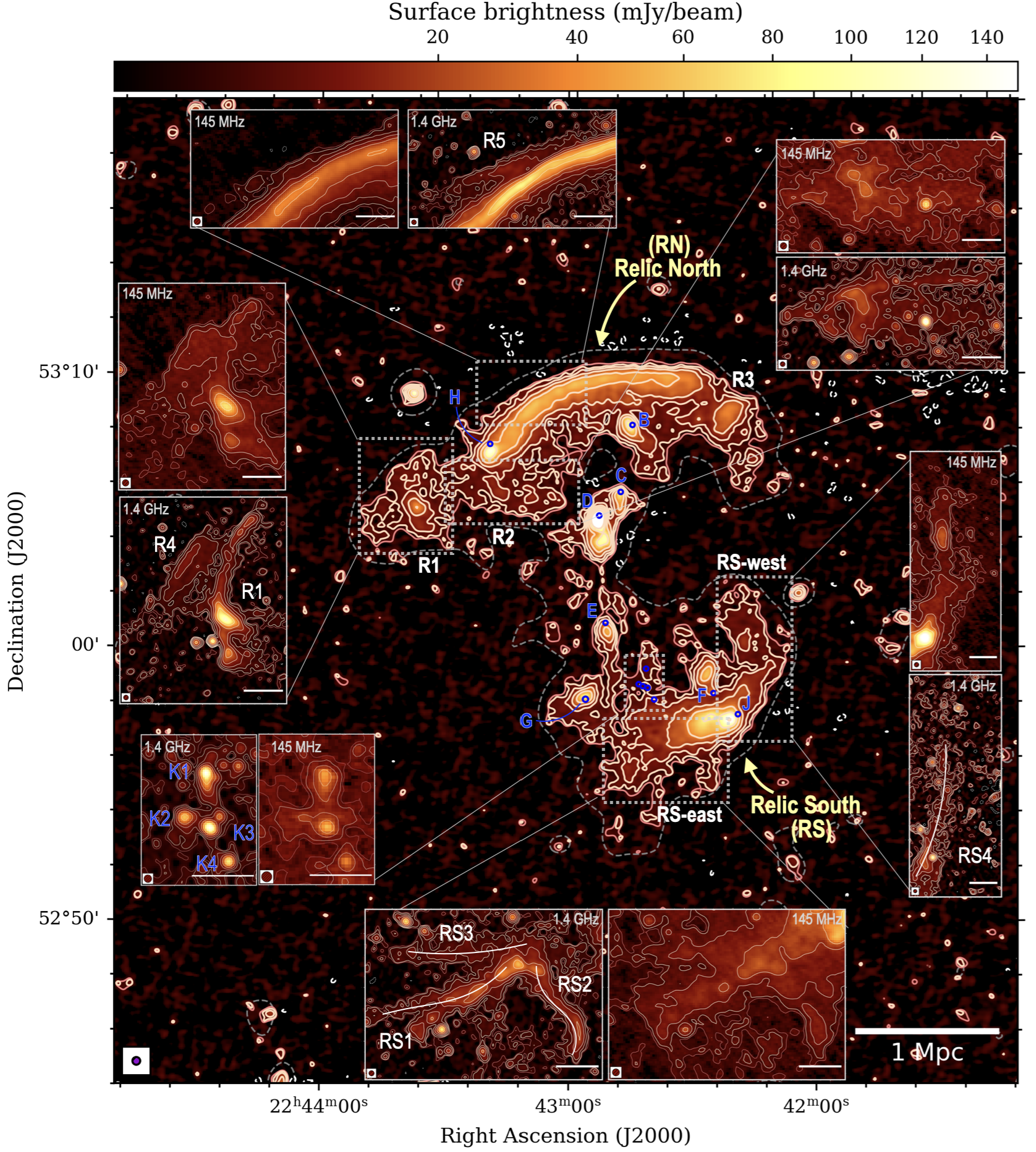}
    \caption{
    Image of \CIZA at $\rm 45\,MHz$, shown at the LBA nominal resolution ($\rm 15\arcsec$ beam). The contour levels are $\rm [-3,2,3,6,12,24,48,96]\times\sigma_{rms}^{45}$, where $\rm \sigma_{rms}^{45}=1.5\,mJy/beam$, with additional dashed $\rm 3\sigma_{rms}$ level from \reviewfirst{$\rm 45\arcsec$ image (\cref{fig:final_images}, right)}.
    Insets show zoomed-in views at different frequencies ($\rm 1.4\, GHz$, \citealt{DiGennaro2018} and $\rm 145\, MHz$, \citealt{Hoang2017}) and comparable resolutions ($6\arcsec$ and $7.5\arcsec$, respectively) of the main substructures of the galaxy cluster. Contour levels in all panels are drawn at $\rm [-3,2,3,6,12,24,48,96]\times\sigma_{rms}$, with $\rm\sigma_{rms}^{1.4}=3.05\,\mu Jy/beam$ and $\rm\sigma_{rms}^{145}=0.11 \,mJy/beam$. The scale bar at the bottom right of each sub-plot corresponds to $\rm 150\, kpc$.
    Source labels are adapted from previous studies \cite[e.g.][]{Stroe2013Discovery, Hoang2017, DiGennaro2018}.
    }
    \label{fig:main_fig}
\end{figure*}

\begin{figure*}
    \includegraphics[width=\textwidth]{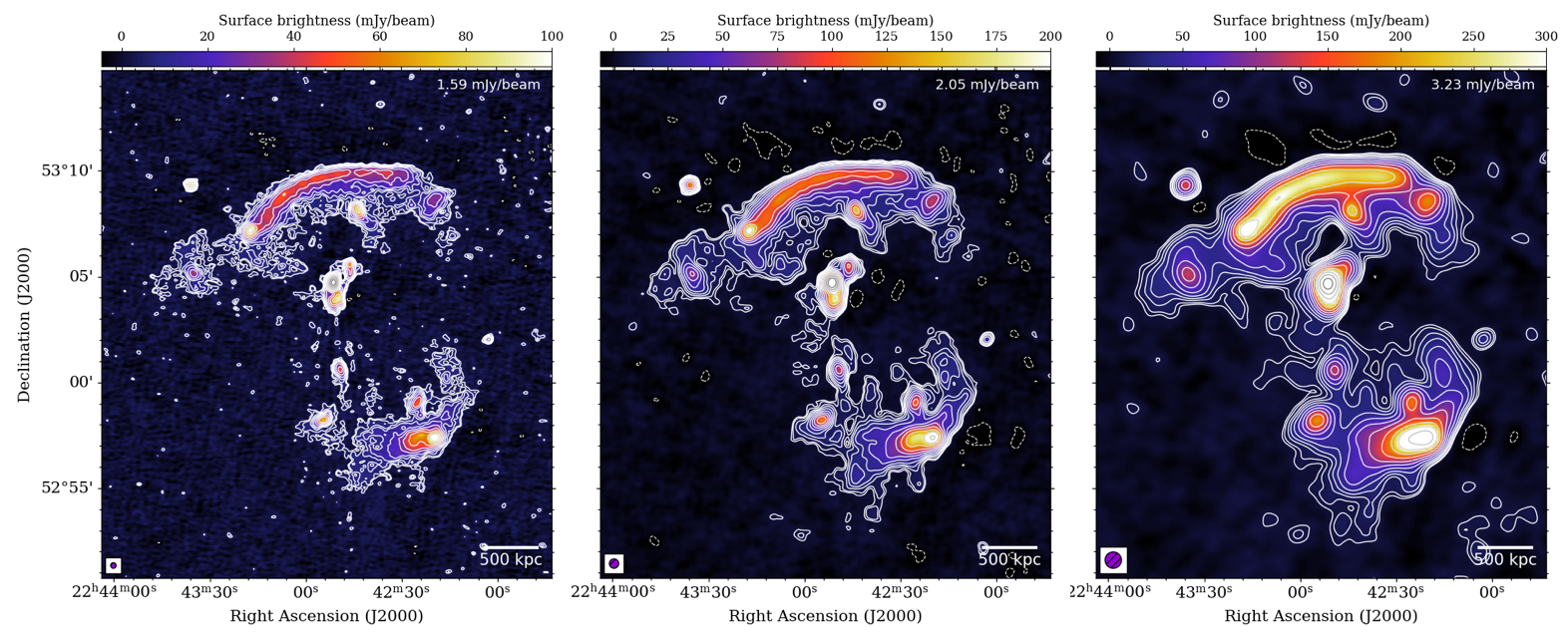}
    \caption{\reviewfirst{Final images at different resolutions. Left: $15\arcsec-$beam, $\rm \sigma_{rms}=$ $\rm 1.59\,mJy/beam$; 
    Centre: $30\arcsec-$beam, $\rm \sigma_{rms}=$ $\rm 2.05\,mJy/beam$;
    Right: $45\arcsec-$beam, $\rm \sigma_{rms}=$ $\rm 3.23\,mJy/beam$;
    Contours are drawn at [-3,3,...]$\rm \times \sigma_{rms}$ with $\sqrt{2}$ steps.}}
    \label{fig:final_images}
\end{figure*}

The first step consists of subtracting all sources outside the region of interest, that is circle of $20'$ around the Sausage cluster, from the visibilities. For this, we take into account the DD calibration solutions of the sources to be subtracted using the facet-mode prediction in \texttt{WSClean} (v 3.4; \citealt{Offringa2014}). 
We then shift the phase centre of the observation to the centre of the region, and we average the data to a resolution of 0.2 MHz in frequency and 32\,s in time. 
Finally, we perform a few additional self-calibration cycles to further refine the image quality.
The final product shows improvement in terms of the image artifacts and also allows flexible re-imaging because of the smaller dataset size. Using \texttt{WSClean}, we produced the final images at high ($15\arcsec$, \cref{fig:main_fig}) mid ($30\arcsec$) and low ($45\arcsec$) resolution (\cref{fig:final_images}). 
We reached a final rms noise of $1.5\,\rm mJy/beam$, in line with the expected sensitivity of $\rm \sim 1-1.5 \,mJy/beam$ for an 8-hour observation \citep{deGasperin2023}.\\

The data reduction of this cluster is made difficult by the two A-Team sources (CasA, CygA) degrees away from the target centre as well as by two strong radio galaxies (NVSS    J224133+531105 and NVSS J223950+52534, see insets in \cref{fig:demix-results}) in the very proximity of the diffuse radio emission. With our calibration strategy, we reached thermal-noise, showing the capability of the \texttt{LiLF} pipeline.\\

\section{Results}\label{sec:results}

\subsection{Radio morphology}
\label{sec:radio_morphology}

In \cref{fig:main_fig} we show the 45 MHz image at LBA $\rm 15\arcsec$ nominal resolution, together with zoomed-in parts showing relevant sub-regions of the radio emission at different frequencies (1.4 GHz, from \citealt{DiGennaro2018} and 145 MHz from \citealt{Hoang2017}). In line with previous work, we detect the two main relics (north and south, RN and RS), and three other regions of diffuse emission (R1-R3) above the \threesigma  level. 
While the LOFAR LBA system provides excellent sensitivity for detecting large-scale features, its nominal $\rm 15\arcsec$-resolution limits the ability to resolve finer filamentary substructures. To address this, we complement \cref{fig:main_fig} with higher-frequency and high-resolution images.

The northern relic (RN) shows the famous sausage-like morphology, with a projected linear size of $\rm\sim 2\, Mpc\times 450\,kpc$ based on the \threesigma of the $15\arcsec$-resolution image (\cref{fig:final_images}, left). It increases to a linear size of $\rm\sim 2.2\, Mpc\times 760\,kpc$ when measured from the $30\arcsec$-resolution image (\cref{fig:final_images}, centre).

Blended with the source RN itself, at its eastern end, we find the double-radio source H, with one jet engulfed in the downstream region of RN and the other which gives rise to the radial tail-like emission in front of the relic. 
Further east, about $\rm 630\, kpc$ from source H, we find an extended, patchy source (R1), with a north-south extension of 570 kpc. R1 displays an irregular arc-like morphology and has a bright peak in the southern part.
South from source H and tightly connected to RN, lies another radial-like  source (R2), which extends for 760 kpc. 
We note that, unlike in higher-frequency observations where these substructures appear fragmented and disconnected from each other —even when compared at similar resolution—, R1, R2 and RN appear interconnected by a homogeneous low-brightness emission that envelops and links these features.
Moreover, in studies at higher-resolution and higher-frequencies  \citep{DiGennaro2018, Raja2024}, both, R1 and R2 sources exhibit substructures and appear to consist of multiple components. Specifically, R1 is further resolved into R4 and R1 (see left panels of \cref{fig:main_fig}); while R2 shows arms-like structures that follow the radial direction.
While the western half of RN shows a sharp edge, which closely coincides with the bright shock rim, we detect emission along the complete length of the eastern half, for a total size of $\rm\sim 700\, kpc$ (see further discussion in \cref{subsec:R5}).
Originally reported by \cite{DiGennaro2018}, and recently confirmed by \cite{Raja2024}, faint diffuse radio emission, labelled as R5, is located north of the eastern edge of RN.
This feature is more clearly defined at higher frequencies and appears more diffuse at LOFAR frequencies (see top panels of \cref{fig:main_fig}).
On the western side of RN, we detect a patchy box-like source labelled as R3. It is elongated towards the radial direction, with a size of $\rm 380\,kpc$. The source R3 is blended with the main relic and further extends towards the southern direction by up to $\rm 430\,kpc$ and it is also detected in \cite{Raja2024} work as R3S.

The southern relic (RS) exhibits an irregular morphology, far thicker and much more contaminated by radio galaxies than the northern one. 
Moreover, its morphology changes significantly with frequency, showing filamentary arms at high frequencies (see RS1, RS2, RS3, RS4 in the bottom and left sub-panels of \cref{fig:main_fig}) while remaining smoother and more diffuse at lower frequencies.
Overall, RS extends for $\rm\sim 1.5\, Mpc\times 520\,kpc$ (linear size based on the \threesigma-contours of the $15\arcsec$-resolution image (\cref{fig:final_images}, left) and $\rm\sim 1.9\, Mpc\times 900\,kpc$ when in the $30\arcsec$-resolution image (\cref{fig:final_images}, centre).

For all the sources (R1, R2, RS1, RS2, RS3, RS4), we find an overall increase in the degree of patchiness with frequency, based on visual inspection. While for the 45 MHz data, part of this effect may be attributed to the relatively low $15\arcsec$ resolution achievable, which smooths out substructures with characteristic widths smaller than this scale, this trend also persists when comparing the morphology at similar $\sim 7\arcsec$ resolution (see insets at 145 and 1400 MHz in \cref{fig:main_fig}). 
Previous studies have shown that a turbulent pre-shock medium can naturally reproduce the small-scale substructures in relics \citep{Dominguez-Fernandez2021}.
Interestingly, recent simulations by \cite{Dominguez-Fernandez2024} showed that an increase in patchiness at higher frequencies would mainly depend on the Mach number distribution of the shock. Specifically, they find that a Mach number distribution with a spread greater than 0.3–0.4 is required to produce observational differences between low- and high-frequencies emission at $\rm 5\arcsec$.

\CIZA hosts a variety of radio sources, many of which have a head-tail morphology and show interactions with the extended emission (see labelled radio galaxies in \cref{fig:main_fig}).
Radio sources, such as sources H and B, are found close to source RN, while radio sources J and F are connected to RS. We find sources C, D, E, and G between the two relics. 
The proximity of these radio galaxies to the diffuse emission regions may play a significant role in shaping relic morphologies and contributing to the overall cluster extended radio emission. In fact, radio galaxies are possible sources of seed electrons dispersed into the ICM \citep{vanWeeren2017NatAs}, which can be re-energized by propagating shocks and turbulence commonly present in galaxy clusters \citep{Brunetti&Jones2014,Vazza2024Galax}.
The role of radio galaxies hosted in \CIZA is discussed throughout this paper when relevant, while more details about them and their optical counterparts can be found in \cite{Stroe2013Discovery, DiGennaro2018}.

\subsection{Integrated spectral index}
\label{sec:int_spec_ind}
\begin{figure}
    \includegraphics[width=\columnwidth]{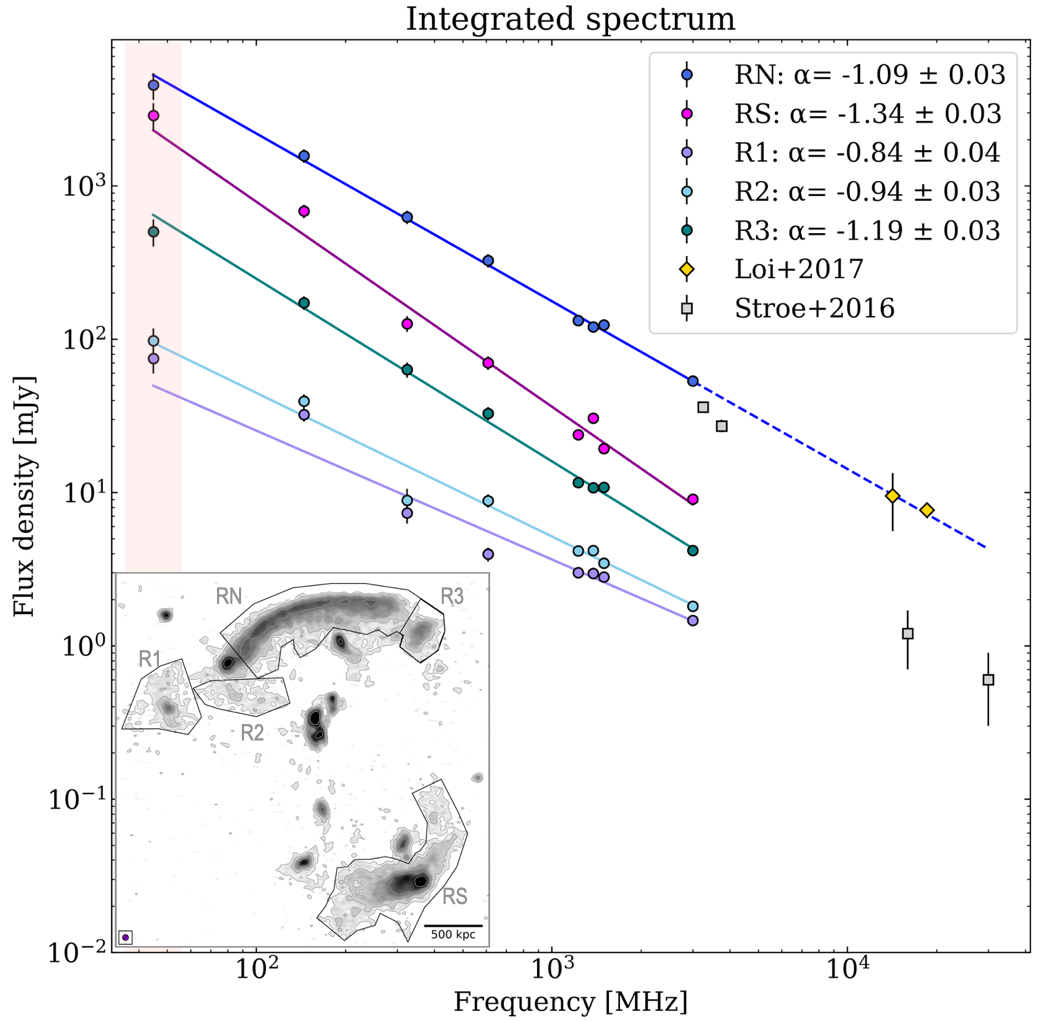}
    \caption{Integrated radio spectrum of RN, RS, R1, R2, and R2 from 45 MHz to 3 GHz. Extraction regions are shown in the bottom left corner. Values obtained in this work at 45 MHz are highlighted in red, while extra points are at 145, 323, 608, 1230, 1380 MHz \citep{Hoang2017}, and 1.5, 3 GHz (\cite{DiGennaro2018}). Above 3 GHz values from high frequencies from \cite{Stroe2016} (grey squares) and \cite{Loi2017} (yellow diamonds) studies are reported.}
    \label{fig:int_specttrum}
\end{figure}

\begin{table*}[h!]
\centering
\caption{Flux densities for the radio relics in \CIZA.}
\label{tab:flux_densities}
\begin{tabular}{cccccc}
\toprule
Freq & RN  & RS  & R1 & R2 & R3 \\
MHz & mJy & mJy& mJy & mJy & R3  \\
\midrule
45   & $4535 \pm 907$  & $2886 \pm 578$  & $449 \pm 91$  & $293 \pm 60$  & $502 \pm 101$ \\ \addlinespace[0.5em]
145  & $1570 \pm 157$  & $684 \pm 68$    & $194 \pm 20$  & $119 \pm 12$  & $173 \pm 17$  \\ \addlinespace[0.5em]
325  & $625 \pm 63$    & $126 \pm 15$    & $44.2 \pm 6.6$  & $26.7 \pm 4.9$  & $63.2 \pm 7.2$    \\ \addlinespace[0.5em]
610  & $326 \pm 33$    & $70 \pm 7$  & $4 \pm 2$  & $27 \pm 3$  & $32.8 \pm 3.3$    \\ \addlinespace[0.5em]
1230 & $132 \pm 7$     & $23.9 \pm 1.4$  & $18.0 \pm 1.0$  & $12.5 \pm 0.7$  & $11.6 \pm 0.7$    \\ \addlinespace[0.5em]
1382 & $120 \pm 6$     & $30.5 \pm 1.6$  & $17.8 \pm 0.9$  & $12.5 \pm 0.6$  & $10.8 \pm 0.6$    \\ \addlinespace[0.5em]
1500 & $124 \pm 6$     & $19.3 \pm 1.1$  & $16.9 \pm 0.9$  & $10.4 \pm 0.6$  & $10.8 \pm 0.6$    \\ \addlinespace[0.5em]
3000 & $53.3 \pm 2.7$  & $9.0 \pm 0.5$   & $8.8 \pm 0.5$   & $5.4 \pm 0.3$   & $4.2 \pm 0.2$     \\ \addlinespace[0.5em]
\bottomrule
\end{tabular}
\begin{tablenotes}
    \item[*] \reviewsecond { The integrated fluxes were measured from $15\arcsec$ images described in \cref{sec:int_spec_ind}. Extraction regions and integrated spectra are shown in \cref{fig:int_specttrum}. }
\end{tablenotes}
\end{table*}

Using 45 MHz LOFAR observations together with additional frequencies from \cite{Hoang2017} (LOFAR HBA: 145 MHz), \cite{Stroe2013Discovery} (GMRT: 325), \cite{vanWeeren2010} (GMRT: 610 MHz; WSRT: 1230, 1382 MHz) and \cite{DiGennaro2018} (JVLA: 1.5, 3 GHz), 
we measured the volume-integrated spectral indices for the main extended sources: RN, RS, R1, R2 and R3.  
In order to minimize biases in measured emission - and resulting spectral shapes - due to differences in the interferometers used to acquire the data and their respective uv-coverages, we used images with a common $\rm 0.2\,k\lambda$ (corresponding to 0.3 deg) inner uv-cut and uniform weighting scheme. 
The integrated flux densities for all sources were calculated within regions defined by the \threesigma-contours of the LOFAR LBA $15\arcsec$ image (see inset of \cref{fig:int_specttrum}).
The RN area is large enough to capture diffuse emission in the downstream region of the shock while avoiding contamination from radio galaxies (such as sources H and B).
Being more extended and irregular, RS embeds a few point sources that were masked during the flux extraction procedure. We note that while the majority of them are compact and well-defined, source J appears to have a tail that steepens and becomes well-mixed with the relic emission  (see \cref{fig:spectral_index_map}), potentially playing a role in the formation of RS itself and affecting its morphology. As no clear distinction between the two sources could be made, we decided to not subtract it. Therefore, we point out that the overall RS spectrum might be artificially slightly steepened because of that.
The spectral index is obtained by a weighted least-square fitting a single power-law function.
The integrated spectra of both relics follow a close power law with a slope of $\alpha_\mathrm{int}^N = -1.09 \pm 0.03$ for RN and $\alpha_\mathrm{int}^S = -1.34 \pm 0.03$ for RS (see \cref{fig:int_specttrum}).
The power-law spectrum obtained for both relics is consistent with the standard scenario for the relic formation, where DSA acceleration occurs from the thermal pool electrons.
These integrated spectral index values are in line with previous studies (see \cref{tab:RNRS_alpha_int}) and are in agreement with other relics studied across a wide frequency range \citep[e.g.][]{Rajpurohit2020_Toothbrush}.
At higher frequencies, a discrepancy exists between \cite{Stroe2016}, who found evidence of spectral steepening for $\rm \nu > 2.5\, GHz$, and a subsequent study by \cite{Loi2017}, which reported no steepening up to 18.6 GHz, attributing the difference to missing diffuse flux in interferometric data. Our new values, focused on the low-frequency end of the spectrum, reveal a spectral behaviour still in line the high-frequency points of \cite{Loi2017} (see \cref{fig:int_specttrum}).
We notice a significant scatter around the best-fit lines for sources R1 and R2 (compared with RN RS and R3). This may result from the low signal-to-noise ratio of the detections and the challenges associated with accurately imaging large, faint and diffuse sources.
However, the resulting best-fit spectral index values ($\alpha_{int}^{R1}=-0.84\pm0.04 $, $\alpha_{int}^{R2}=-0.94\pm0.03$) are in agreement with higher-frequency studies, such as \cite{Stroe2013Discovery, Hoang2017, Raja2024}.
The variety of spectral indices highlights the complexity of the merger, which cannot be simply attributed to the emission from a primary shock and its counter-shock, and  suggests the presence of a combination of processes, perhaps including multiple shocks on different scales.
If these substructures are interpreted as additional relic-like regions or smaller-scale shock structures, their relatively flatter integrated spectral indices would imply strong Mach numbers, suggesting the presence of more energetic shocks (see \cref{sec:mach_number} for further details).

\begin{table}[h!]
\centering
\caption{Integrated spectral index estimates for the northern (RN) and southern (RS) radio relics in \CIZA.}
\label{tab:RNRS_alpha_int}
\resizebox{\columnwidth}{!}{ 
\begin{tabular}{cccc}
\toprule
 & \multicolumn{2}{c}{Source} & \\
\cmidrule(r){2-3}
 & RN & RS & Ref. \\ 
\midrule
$\alpha_{0.153}^{2.7}$  & $-1.06 \pm0.04$ & $-1.29\pm0.04$   & \cite{Stroe2013Discovery} \\ \addlinespace[0.5em]
$\alpha_{0.145}^{2.3}$  & $-1.11 \pm0.04$ & $-1.41\pm0.05$   & \cite{Hoang2017} \\ \addlinespace[0.5em] 
$\alpha_{0.145}^{18.6}$ & $-1.12 \pm 0.03$ & -               & \cite{Loi2020}   \\  \addlinespace[0.5em] 
$\alpha_{1.5}^{3}$      & $-1.19 \pm 0.05$ & $-1.12\pm0.07$  & \cite{DiGennaro2018} \\  \addlinespace[0.5em] 
$\alpha_{0.045}^{2.3}$  & $-1.09 \pm 0.03$ & $-1.34\pm0.04$  & this work \\  \addlinespace[0.5em] 
\bottomrule
\end{tabular}
}
\vspace{0.5em}
\end{table}

\subsection{Spectral index maps}
\label{sec:spec_ind_maps}
\begin{figure*}
    \includegraphics[width=\textwidth]{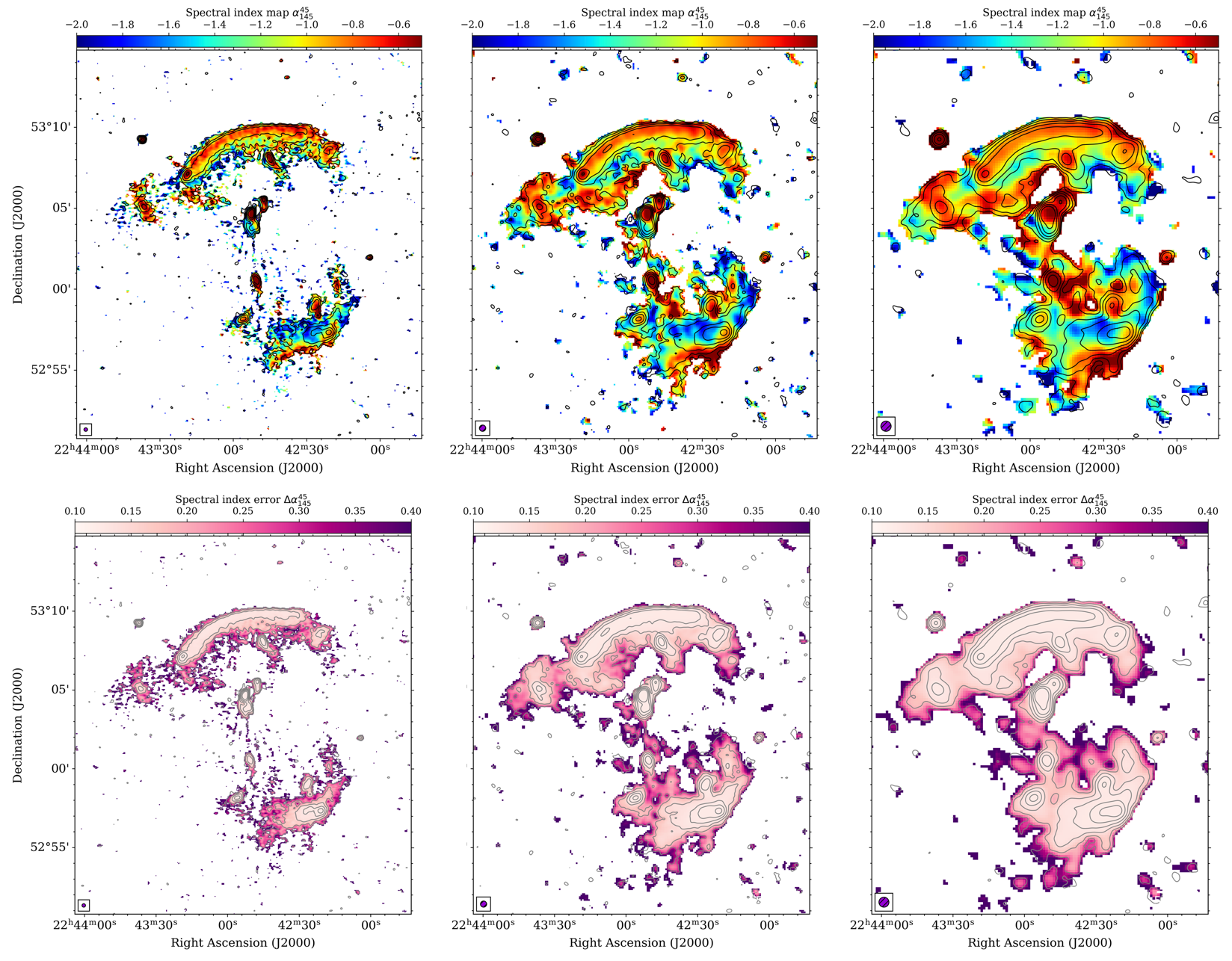}
    \caption{Top panel: Spectral index map between 45 and 145 MHz at a spatial resolution of $ 13\arcsec\times 13\arcsec$ (left), $ 23\arcsec\times 23\arcsec$ (middle) and $ 37\arcsec\times 37\arcsec$ (right).
    Bottom panel: Relative spectral index error $\Delta \alpha_{\rm 145~MHz}^{\rm 45~MHz}$. 
    Pixels with surface brightness values below $2\sigma_{\rm rms}$ in the two images were blanked and $[3,6,12,24,48,96]\times \sigma_{\rm rms}^{\rm 45~MHz}$ contours are over-plotted.}
    \label{fig:spectral_index_map}
\end{figure*}

We produced spectral index maps between 53~MHz and 144~MHz (\cref{fig:spectral_index_map}). To this end, we re-imaged LBA and HBA data, using Briggs weighting with Robust parameter $R=-0.25$, common $uv$-range ($\rm 80 \, \lambda$ inner $uv$-cut, corresponding to $0.7\deg$) and convolved to the common beam size. 
In each map, we measured spectral indices for pixels with a flux density $\geq 2\sigma_{\rm rms}$ in both frequencies. \\

Both relics exhibit a spectral index gradient that decreases from the outskirts towards the cluster centre.
In the $13\arcsec$-resolution map (\cref{fig:spectral_index_map}, left), the RN shows a spectral index gradient from the outskirts towards the centre of the cluster, \reviewfirst{from $\alpha_{45
}^{145} = -0.7\pm0.15$ to $ -1.8\pm0.35$; while the RS goes from $-0.5\pm0.3$ to $ -2\pm0.2$ across its width.}
In the mid-resolution $23\arcsec$-map (\cref{fig:spectral_index_map}, centre) the northern relic shows the spectral index gradient followed by a re-flattening. \reviewfirst{These flat $\alpha_{45
}^{145} = -0.7\pm0.2$ values correspond} to sources I and R2, two radial structures in the downstream region of the shock (see \cref{fig:final_images}).
The patchy source R3 appears to have a similar gradient to the main relic RN, but appears more diffuse and less pronounced.
While morphologically we could not detect a clear distinction in R1 (\cref{sec:radio_morphology}), the spectral index maps allow us to identify two areas. R1-east, with values $-0.8\pm0.1$, and R1-west (spatially corresponding to R4) show more values around $-1.3\pm0.2$. 
This clear difference in terms of the spectral index implies that there are two distinct sources.
Source R2 shows fluctuations of $\alpha_{45}^{145}$ from $-0.8$ to $-1.2$ and no clear gradient or trend. This patchy nature in terms of morphology and spectral index (both in this work and at a higher frequency - see Fig. 4 in \cite{DiGennaro2018}) might indicate a scenario where this relic emission is seen with a larger inclination, that is more face-on. This behaviour with mixed values of spectral index is for example seen in Abell 2256 \citep{Rajpurohit2022_A2256}, one of the most famous face-on radio relic.

The southern relic, RS, shows an irregular morphology, which is reflected in the variance of spectral indices and their spatial distribution. In fact, despite showing a gradient reaching even to steeper values ($\alpha_{45}^{145}\sim -2$) with respect to RN, only the south-east part of the relic shows the expected flat rim followed by a decrease. On the other hand, the western part of RS shows already steep values at the furthest edge ($\sim -1.8\pm0.2$) and no inward steepening. These broad steep-spectrum areas explain why RS fade quickly at higher frequencies ($\rm \nu\gtrsim 600\, MHz$), where primarily the southern edge remains bright.

\subsection{Curvature maps}
\label{sec: curvature_maps}

The LOFAR LBA 45 MHz data represents the lowest frequency observation available for this target and it offers a unique opportunity to study the spectral curvature across an unprecedentedly wide frequency range. 

\begin{figure*}
    \includegraphics[width=\textwidth]{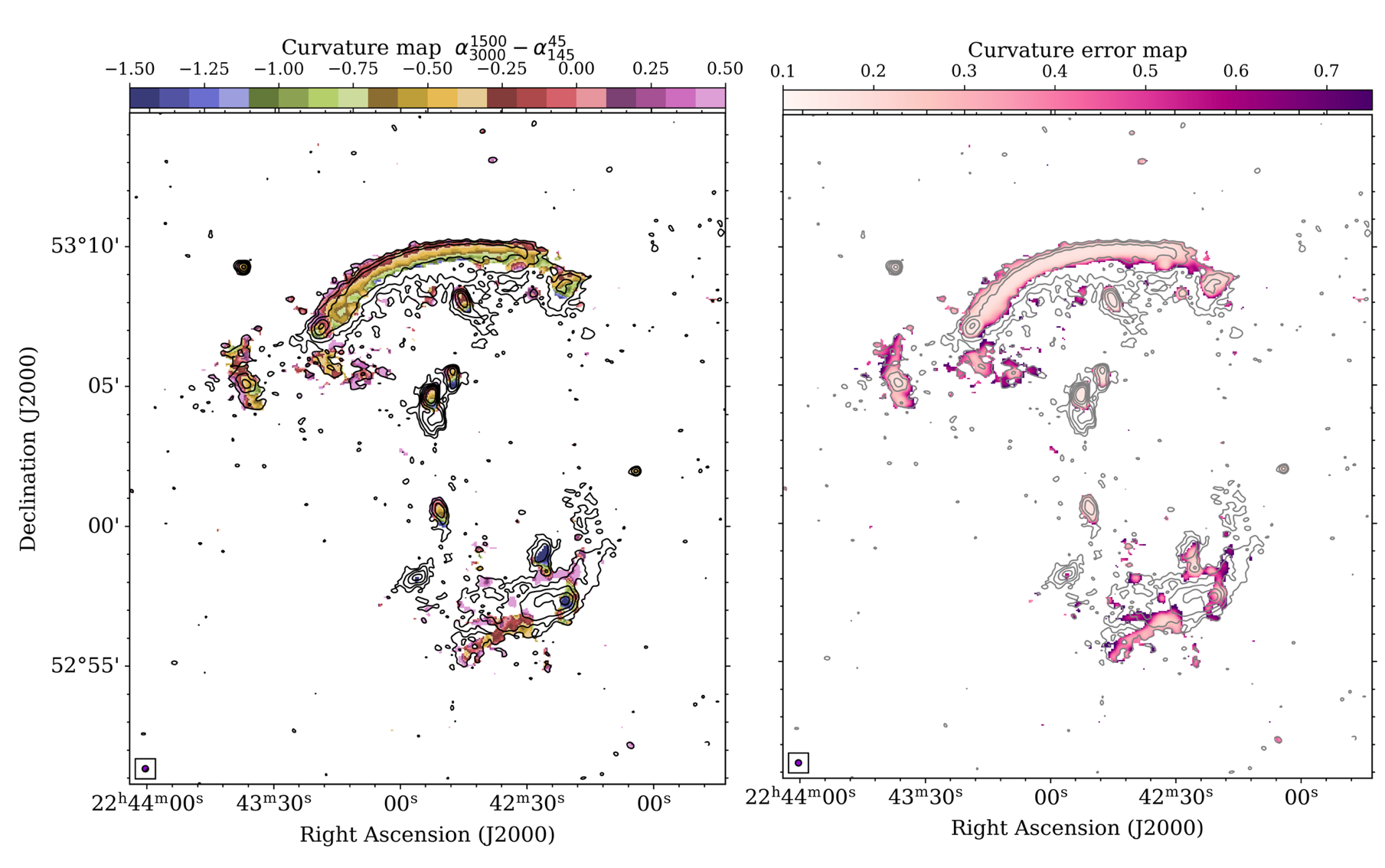}
    \caption{Four-frequency (45, 145, 1500, 3000 MHz) spectral curvature map. %
    Pixels with surface brightness values below $2\sigma_{\rm rms}$ in the two images were blanked and $[3,6,12,24,48,96]\times \sigma_{\rm rms}^{\rm 45~MHz}$ contours are over-plotted.}
    \label{fig:curvature_map}
\end{figure*}

We produced curvature maps using images at 45, 145, 1500 and 3000 MHz convolved to the same $15\arcsec$ resolution. As discussed in \cref{sec:spec_ind_maps}, we only considered pixels above the $\rm 2\sigma$ threshold.
In agreement with previous work \citep{Stroe2013Discovery, DiGennaro2018}, we define the four-frequency spectral curvature as

\begin{equation}
\label{eq:C}
    C = \alpha_{\rm high} - \alpha_{\rm low} ,
\end{equation}
where $\rm \alpha_{high}=\alpha^{1500}_{3000}$ and $\rm \alpha_{low}=\alpha^{45}_{145}$.
\cref{eq:C} is consistent with the three-frequency definition used \cite{LeahyRoger1998}. In this framework, $\rm C<0$ for $ \rm |\alpha_{high}|>|\alpha_{low}|$, which is the case of a concave spectrum affected by standard spectral ageing. Conversely, $C<0$ corresponds to an inverted spectrum, where $\rm |\alpha_{high}|<|\alpha_{low}|$, that is the spectrum is steeper at lower frequencies.
The resulting spectral curvature maps are shown in \cref{fig:curvature_map}. 
The curvature at source RN increases from $C=0$ in the outer part to values of $\sim -1.5$ in the downstream area. This is expected from the DSA theory which predicts a power-law energy distribution at the front of the travelling shock, where the electron population is freshly accelerated, leading to a relatively flat and constant spectral index across frequencies.
The ranges of values obtained by this analysis agree with the previous curvature analysis by \cite{Stroe2013Discovery}. However, while \cite{Stroe2013Discovery} reported small-scale variations along the source with a size of $\rm \sim 64~kpc$, we observe fewer pronounced variations in our data. Along the shock rim, the spectral curvature appears  patchy and discontinuous.
R1 shows a slight change in curvature from $\sim 0.2$ to $\sim-1.25$ curvature.  However, the large uncertainties at the edges $(\Delta C \sim 0.50)$ prevent meaningful conclusions about this gradient, as within these uncertainties, the curvature values remain consistent with a power-law spectrum.
In the same way, also source R2 shows a rather small curvature with values  $\sim - 0.3$.
Source R3 shows a gradient similar to the one of the main relic, going from $\sim -0.45$ to $\sim-1.25$.
The southern relic, RS, does not show areas with zero curvature with constant values of $C\sim -0.2$ to $-0.5$ in the RS-south area. 
Finally, point sources such as B, C, D, E, F show a canonical aged spectrum with increasing curvature the furthest away from the AGN reaching values of $C\sim-1.5$ at the ends of the jets.

\section{Discussion}\label{sec:discussion}

\begin{table}[h!]
\centering
\caption{Mach number estimates for the northern (RN) and southern (RS) radio relics in \CIZA.}
\label{tab:RNRS_Mach_numbers}
\resizebox{\columnwidth}{!}{ 
\begin{tabular}{lccr}
\toprule
 & \multicolumn{2}{c}{Source} & \\
\cmidrule(r){2-3}
 & RN & RS & Ref. \\ 
\midrule
$\mathcal{M}_{\rm R}^{\star}$       & $4.6^{+1.3}_{-0.9} $           & $ -$   & \cite{vanWeeren2011GMRT}  \\ \addlinespace[0.5em]
$\mathcal{M}_{\rm R}^{\star}$       & $4.58\pm1.09$ & $2.81\pm0.19$  & \cite{Stroe2013Discovery}   \\ \addlinespace[0.5em]
$\mathcal{M}_{\rm R}^{\star}$       & $4.4^{+1.1}_{-0.6}$   & $2.4\pm0.1$           & \cite{Hoang2017}      \\  \addlinespace[0.5em] 
$\mathcal{M}_{\rm R}^{\star}$       & $4.2^{+0.4}_{-0.6}$   & $-$  & \cite{Loi2020}   \\ \addlinespace[0.5em]
$\mathcal{M}_{\rm R}^{\star}$       & $4.8\pm0.8$ & $2.6\pm0.1$  & this work  \\ \addlinespace[0.8em]
$\mathcal{M}_{\rm R}^{\cross}$       & $2.90^{+0.10}_{-0.13}$ & $-$  & \cite{Stroe2014_Spectral_age_modelling}   \\ \addlinespace[0.8em]
$\mathcal{M}_{\rm R}^{\bullet}$       & $2.9^{+0.3}_{-0.2}$   & $2.2^{+2.1}_{-0.4}$           & \cite{Raja2024}      \\  \addlinespace[0.5em] 
$\mathcal{M}_{\rm R}^{\bullet}$       & $2.58 \pm 0.17$       & $2.10 \pm 0.08$        & \cite{DiGennaro2018} \\  \addlinespace[0.5em] 
$\mathcal{M}_{\rm R}^{\bullet}$       & $2.7^{+0.6}_{-0.3}$   & $1.9^{+0.3}_{-0.2}$           & \cite{Hoang2017}      \\  \addlinespace[0.5em] 
$\mathcal{M}_{\rm R}^{\bullet}$       & $2.9\pm0.4 $          & $2.9\pm0.8$   & this work  \\ \addlinespace[1em] 
$\mathcal{M}_{\rm X}$   & $2.7^{+0.7}_{-0.4}$      &  $1.7^{+0.4}_{-0.3}$    & \cite{Akamatsu2015} \\  \addlinespace[0.5em]
$\mathcal{M}_{\rm X}$   & $2.54^{+0.64}_{-0.43}$   & -      & \cite{Ogrean2014Chandra}\\  \addlinespace[0.5em]
$\mathcal{M}_{\rm X}$   & $3.15\pm0.52$            & -      & \cite{Akamatsu2013} \\  \addlinespace[0.5em] 
$\mathcal{M}_{\rm X}$   & -                        & $1.2-1.3$      & \cite{Ogrean2013XMM}\\  \addlinespace[0.5em] 
\bottomrule
\end{tabular}
}
\vspace{0.5em}
\small
$^{\star}$: derived from volume-integrated spectral index via \cref{eq:alpha_int_inj};\\
$^{\cross}$: derived from spectral ageing modelling;\\
$^{\bullet}$: derived from local spectral index measurement at the relic edge from images or maps.
\end{table}

\subsection{Mach number from radio data}
\label{sec:mach_number}

Based on the assumption of standard DSA theory, the shock's Mach number, $\mathcal{M}$, can be estimated from the radio spectrum via \citep{Drury1983, Blandford1987}
\begin{equation}
\label{eq:M_delta_inj}
    \delta_{\rm inj} = 2 \frac{\mathcal{M}^2+1}{\mathcal{M}^2-1} ,
\end{equation}
where $\delta_{\rm inj}$ is the index of the electrons power-law energy distribution injected by the shock ($\rm dN(p)/dp \propto p^{-\delta_{inj}}$). This is related to the injection spectral index as $\rm \alpha_{inj} = -(\delta_{inj}-1)/2$. 
The volume-integrated spectral index on a radio relic, where there is a balance between acceleration and energy losses, is related to the injection index as \citep{Kardashev1962}
\begin{equation}
\label{eq:alpha_int_inj}
    \alpha_{\rm int} = \alpha_{\rm inj}+0.5 .
\end{equation}
Thus \cref{eq:M_delta_inj} can be re-written as 
\begin{equation}
\label{eq:M_alpha_inj}
    \mathcal{M}  = \sqrt{\frac{2 \alpha_\mathrm{inj} - 3}{2 \alpha_\mathrm{inj} + 1}} ,
\end{equation}
so that $\mathcal{M}$ can be derived directly from radio flux density measurements.

From integrated spectral index values of \cref{sec:int_spec_ind}, we derived injection indices between 45 and 145 MHz of $\alpha_\mathrm{inj}^N = -0.59 \pm 0.03$ and $\alpha_\mathrm{inj}^S = -0.84 \pm 0.03$ and Mach numbers of $\mathcal{M}_N = 4.8\pm0.8$ and $\mathcal{M}_S = 2.6 \pm 0.1$. 

These results are consistent with previous work (see also \cref{tab:RNRS_Mach_numbers}): \cite{vanWeeren2010} found $\mathcal{M}_N=4.6^{+1.3}_{-0.9}$, from spectral index map, \cite{Stroe2013Discovery} measured $\mathcal{M}_N = 4.58 \pm 1.09$ and $\mathcal{M}_N = 4.58 \pm 1.09$, from spectral index maps and colour-colour analysis, \cite{Hoang2017} ($\mathcal{M}_N =4.4^{+1.1}_{-0.6}$) and \cite{Loi2020} ($\mathcal{M}_N =4.2^{+0.4}_{-0.6}$) derived it from the integrated spectrum. Although consistent with one another, these estimates are considerably higher than the Mach numbers derived from X-ray data, reflecting the known trend $\rm \mathcal{M}_{R}>\mathcal{M}_{X-ray}$, as also expected from simulations \citep{Wittor2021, Whittingham2024}.
Deriving the Mach number from integrated spectral properties relies on the assumption of constant shock properties (such as velocity, compression ratio and magnetic field) and that the downstream ageing gas maintains steady conditions over time, with a balance between particle injection, acceleration, and radiative losses. 
However, several studies \citep{Wittor2019,Dom-Fern2024,Whittingham2024} using cosmological MHD simulations showed that shocks in radio relics are complex structures, with significant spatial variation in both Mach numbers and magnetic fields.
\\

\begin{figure*}
    \includegraphics[width=\textwidth]{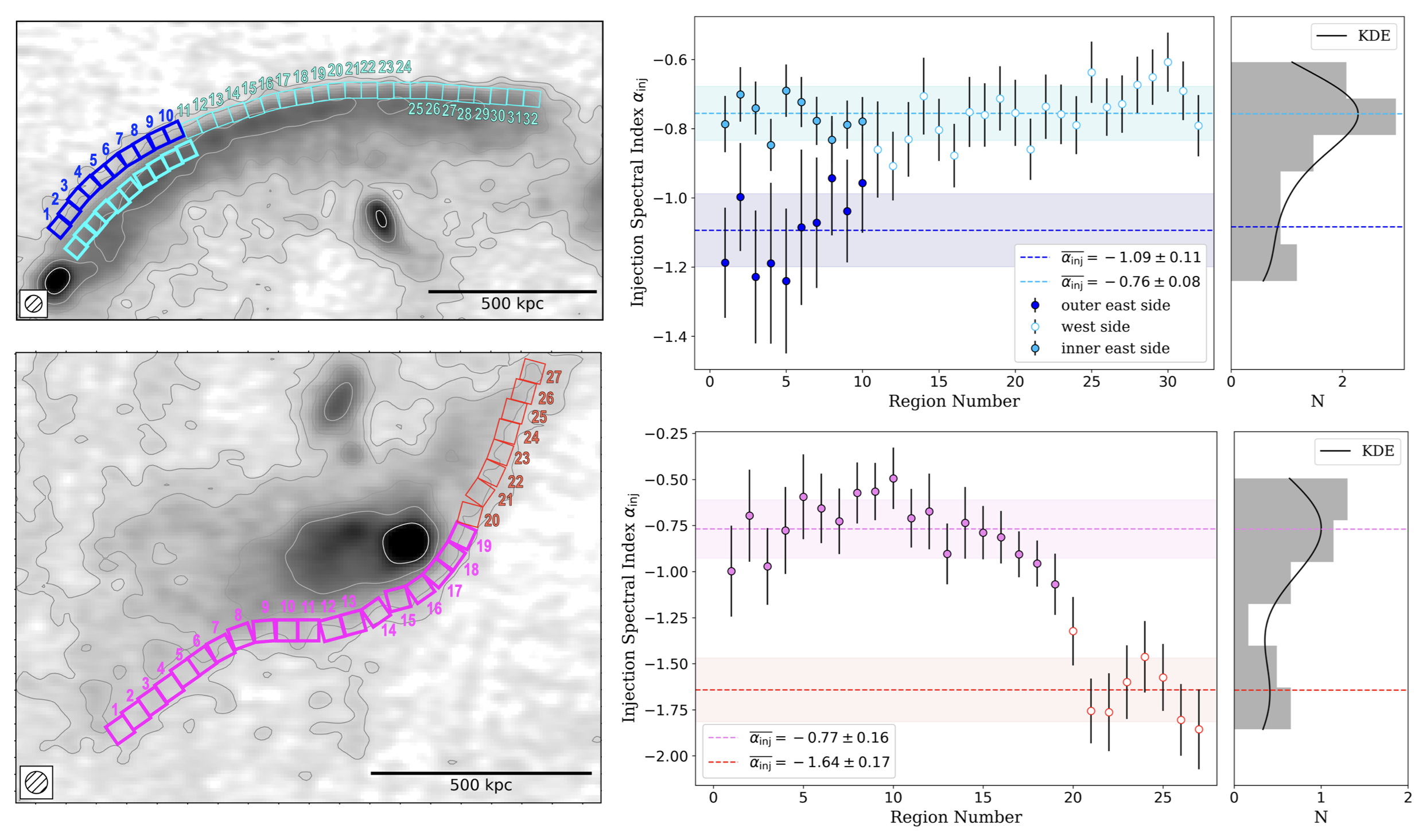}
    \caption{Injection spectral index calculated in $15\arcsec$-box ($\rm \sim48\,kpc$ separation) along the relics' edges. 
    Left: Zoom-in over the RN (top) and RS (bottom). 
    Right: Injection spectral index profiles extracted from 45-145 MHz $15\arcsec$-resolution images. The right panels display  the density-normalized histograms together with a kernel density estimation (KDE) curve for a smooth and continuous representation of the data distribution.
    For each relic, the data are colour-coded by the different groups identified by the GMM. The GMM means are plotted as horizontal dashed lines.}
    \label{fig:spectral_index_injection}
\end{figure*}

A more accurate method to calculate the Mach number from the radio spectral index involves directly measuring the injected spectral index \ainj, by identifying $\rm \alpha$ at the shock front - presumed to be injection region - rather than relying on \aint via \cref{eq:alpha_int_inj} and \cref{eq:M_alpha_inj}.
This is because, at the shock front, where particles have been recently (re-)accelerated, the injection spectral index is expected to be flat, while it steepens downstream due to synchrotron and IC energy losses. Low-frequency observations are particularly well-suited for this analysis, as they are less affected by these energy losses, providing a more reliable proxy for the injection spectral index. 

The highest achievable resolution with LOFAR LBA is $\rm 15\arcsec$. As this sets the minimum size of the regions over which we calculate the flux and corresponding spectral index, we need to ensure that these regions correspond to the physical thickness of the shock, thus minimizing the risk of mixing different electron populations.
Considering a downstream velocity of $\rm \sim 1000\, km/s$ \cite{vanWeeren2010}, the travel time for the electrons to cross a distance equivalent to the $\rm 15\arcsec$ beam-size  is $\rm \sim 50$ Myr. This is about 5 times shorter than the typical electron cooling time ($\rm\sim 360~Myr$, for relativistic electrons in $\rm\sim \mu G$ magnetic field and detected at $\rm45$ MHz). Thus, we can assume that the energy losses due to synchrotron and IC process are negligible over the sampled distance, ensuring the measured spectral index is an accurate measure of the injection spectral index at the shock front.

To measure \ainj, we used the $15\arcsec-$resolution images at $45$ and $145$ MHz as done for spectral maps (\cref{sec:spec_ind_maps}) and estimated the injection spectral index in different beam-size regions along the relic edges (\cref{fig:spectral_index_injection}, left). 
In an ideal scenario with no projection effects and an edge-on view, the shock front coincides with the furthest contour of the relic emission and a common approach is to draw these regions along the furthest $\rm 3\sigma_{rms}$-contour.
However, from the spectral maps of \cref{fig:spectral_index_map}, we note that the furthest edge of the northern relic RN (in particular the eastern half) does not correspond to the brightest rim or the region with the flattest spectral index.
The picture is even more complex for the southern relic RS, due to its very irregular morphology and lack of a well-defined shock-like edge. 
Given these complexities, we first measured the spectral index in regions along the furthest edge of both relics, as this would be the expected location of the shock front (\cref{fig:spectral_index_injection}, left). 
This initial choice was made to avoid bias from prior knowledge of the relics morphology and ensure a systematic approach.
To assess whether the spectral index values across the edges of the relics represent distinct  populations, we employed a Gaussian Mixture Model (GMM) to classify the data into clusters. 
We then used information criteria, such as Akaike Information Criterion (AIC, \cite{Akaike1974}) and Bayesian Information Criterion (BIC, \cite{Schwarz1978}) to validate the GMM results by evaluating model fit.
We applied a 2-sample Kolmogorov-Smirnov (KS, \cite{Massey1951}) test to independently verify that the identified clusters represent statistically distinct populations. The results are shown in (\cref{fig:spectral_index_injection}, right).

For the southern relic, RS, the GMM analysis identified two distinct clusters with mean spectral index values of $ \bar{\alpha}_{\mathrm{inj}} = -0.77\pm 0.16 $ and $ \bar{\alpha}_{\mathrm{inj}} = -1.64\pm 0.17$.
Both the BIC and AIC confirmed that two Gaussian components ($k=2$) best describe the data. The KS test further supported this finding, yielding a test statistic of 1 and a highly significant p-value $1.9\times10^{-7}$. 
The two distinct statistical populations also correspond to two separated locations of the relic (see \cref{fig:spectral_index_injection}): the flatter one corresponds to the south-east section of the relic (RS-east of \cref{fig:main_fig}), while the steeper is the top right edge (RS-west of \cref{fig:main_fig}).
This indicates that the source RS contains two statistically distinct populations of spectral index values along the outer edge, potentially reflecting spatial variations in shock properties or a combination of different effects given the more complex RS dynamics and the potential interaction with source J.
One possible explanation for the spectral difference is the effect of slow compression, acting on a timescale longer than the radiative losses, enacted by source J. Compression generally increases both the particle energy and magnetic field strength, but its impact can differ based on the timescale it happens.
If compression occurs rapidly, the energy gained can effectively accelerate particles before significant losses occur. On the other hand, if compression occurs over a timescale comparable to or longer than the cooling time, it will increase the magnetic field strength, but at the same time, radiative losses ($\rm \propto \gamma^2 B^2$) will become dominant, causing high-energy particles to lose energy faster than they gain from compression.
This might explain why the north-west extension of source RS is drastically reduced at high frequencies (see \cref{fig:main_fig}).

We performed the same analysis also for RN, finding two statistically distinct populations (KS test statistic 1 and p-value $3\times10^{-8}$) with mean spectral index values of $\rm \bar{\alpha}_{\mathrm{inj}} = -0.76\pm 0.08 $ and $\rm \bar{\alpha}_{\mathrm{inj}} = -1.09\pm 0.11 $. 
Our measured $\rm \bar{\alpha}_{\mathrm{inj}}$ is also in perfect agreement with the value obtained by \cite{Stroe2014_Spectral_age_modelling} ($0.77^{0.03}_{0.02}$), which was instead inferred by modelling emission with different ageing models. 
We note that the steepening of the spectral index in the eastern part of RN (blue boxes and points of \cref{fig:spectral_index_injection}) coincides with the source R5, which is located in front of RN. 
However, if we follow the morphology of the main shock - traced by the cyan bold regions and cyan-filled points in \cref{fig:spectral_index_injection} - we find values consistent with those in the western part of RN. Notably, these cyan regions align with the brightest rim of the relic, the location where, based on visual inspection, one would expect the shock front to be.
The R5 region is interesting because, on one hand, it exhibits a  significantly steeper spectral index, with $\rm \bar{\alpha}_{145}^{45} \sim -1.09$; on the other hand, if considered part of the main relic, it affects the overall morphology of RN, broadening its width. The characterization of this region will be discussed in \cref{sec: SB_profiles} and \cref{subsec:R5}.

These results show that even when small projection effects are present (as in the case of \CIZA) the edges of the relic present significant spectral index variation across their length. 
Variation of $\rm \alpha_{inj}$ might imply variations of Mach numbers. This is consistent with previous studies that show how shocks producing radio relics often exhibit Mach number variations across the shock fronts, due to the interaction with a turbulent medium. 
In particular, numerical simulations \citep{Skillman2013,Wittor2016,Dominguez-Fernandez2021} have proven that the distribution of Mach numbers naturally arises when a uniform shock propagates through a turbulent medium.
Observational evidence supporting this scenario was reported by \cite{deGasperin2015}, who found a gradient in the Mach number along the shock front of the radio relic in PSZ1 G108.18-11.53.

Moreover, this demonstrates that estimating \ainj by simply measuring values across the full relic following the \threesigma-contour can result in misleading averages. 
Since the resulting $\rm \bar{\alpha}$ values would appear artificially steeper without distinguishing between different regions, such an approach risks underestimating the shock acceleration efficiency and the corresponding Mach number, possibly leading to oversimplified interpretations of the underlying physics.

\begin{figure}
    \includegraphics[width=\columnwidth]{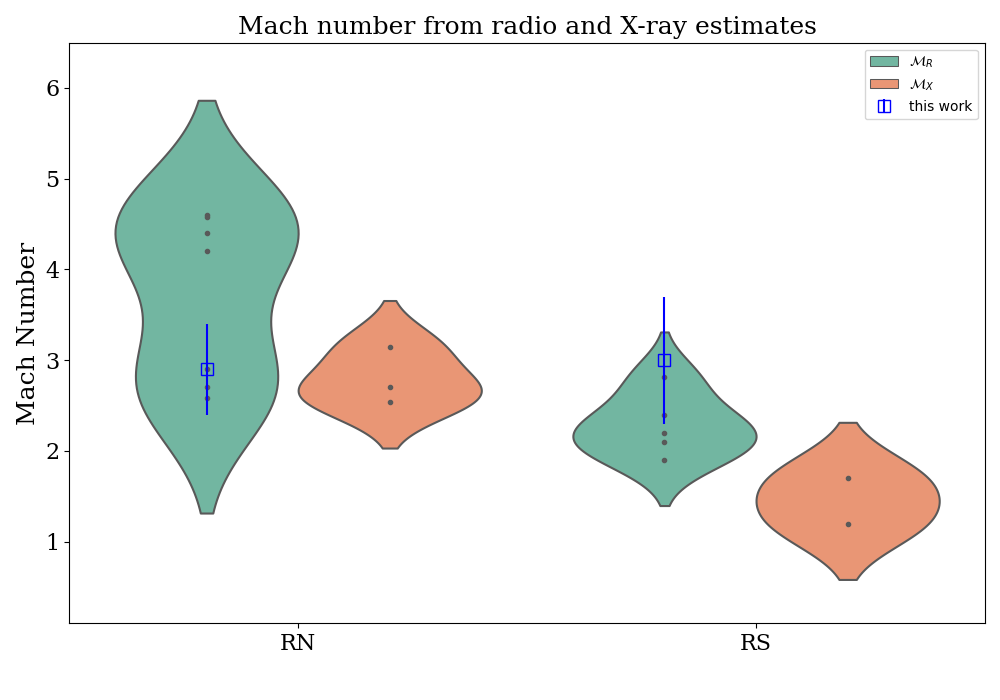}
    \caption{Mach number estimates for northern (RN) and southern (RS) radio shocks in \CIZA, derived from the radio spectral index ($\mathcal{M}_{\rm R}$, teal) and the ICM X-ray analysis ($\mathcal{M}_{\rm X}$, orange). The violin plots illustrate the distribution of Mach number estimates from the literature (listed in \cref{tab:RNRS_Mach_numbers}), highlighting the range and density of measured values. Overlaid points represent individual measurements, while the blue empty squares are the values derived in this work (see \cref{sec:mach_number}).}
    \label{fig:Mach_distribution_plot}
\end{figure}

Considering the results above, we define the injection indices of $\bar{\alpha}_{\mathrm{inj}}^N  = -0.76 \pm 0.08$ and $\bar{\alpha}_{\mathrm{inj}}^S = -0.77 \pm 0.16$. This corresponds to Mach numbers $\mathcal{M}_N = 2.9 \pm 0.4$ and $\mathcal{M}_S = 2.9 \pm 0.8$.
The derived Mach number for RN is consistent with previous literature estimates based on local spectral index measurements (see $\mathcal{M}_R^{\bullet}$ listed in \cref{tab:RNRS_Mach_numbers}). In contrast, the Mach number obtained for RS is significantly higher than in previous studies, though still within agreement given the large associated uncertainty.
So far the Mach numbers derived for \CIZA double-relic system are found to be quite different from each other with $\mathcal{M}_N$ almost double $\mathcal{M}_S$ when measured from the volume-integrated spectral index (see $\mathcal{M}^{\star}$ values in \cref{tab:RNRS_Mach_numbers}). 
The discrepancy is reduced when Mach numbers are derived from $\rm \alpha_{inj}$ at the shock location (see $\mathcal{M}^{\bullet}$ values in \cref{tab:RNRS_Mach_numbers}), yet $\mathcal{M}_N> \mathcal{M}_S$.
The results from our analysis instead, suggest that the two main shocks have comparable strengths. The overall distribution of Mach number estimates from both radio and X-ray studies is visually shown in \cref{fig:Mach_distribution_plot}, together with the newly derived values from this work.
This is not unexpected, as in nearly equal-mass mergers (mass ratio $\sim 1:1$) such as \CIZA, both shocks are often similar in strength, though local variations in pre-shock conditions and magnetic fields can introduce asymmetries.
High consistency in Mach number couples is also observed in other well-studied double relic systems, where both relics have been characterized in detail. That is the case of Abell 1240 \citep[][$\mathcal{M}_1=2.4, \mathcal{M}_2=2.3$]{Hoang2018}, PSZ1 G108.18-11.53 \citep[][$\mathcal{M}_1=2.33, \mathcal{M}_2=2.20$]{deGasperin2015}, Abell 2345 \citep[][$\mathcal{M}_1=2.8, \mathcal{M}_2=2.2$]{Bonafede2009}, and ZwCl 0008.8+5215 \citep[][$\mathcal{M}_1=2.2, \mathcal{M}_2=2.4$]{vanWeeren2011_ZwCl0008}, where the Mach numbers of the opposing relics are found to be very similar.

It is important to note the significant discrepancy between the two methods used to derive the Mach numbers. In the standard DSA framework, we can apply \cref{eq:alpha_int_inj}, \cref{eq:M_alpha_inj}. That means that ideally, the injection spectral index calculated at the shock from $\rm \alpha_{inj}$ should be 0.5 flatter than the integrated value $\rm \alpha_{int}$ derived in \cref{sec:int_spec_ind}. However, for this and other clusters \citep{Botteon2020_shock_acceleration}, this relation does not hold exactly. 
A simple explanation might be related to projection effects. In case of projection, the observed radio emission possibly contain a mixture of emission from different regions along the line of sight. Thus, it becomes difficult to isolate the pure injection spectrum. In this scenario, the measured local $\rm \alpha_{inj}$ will appear steeper than the one obtained via \cref{eq:alpha_int_inj} from the integrated spectrum, due to mixing of emission with slightly different spectral ages.
Alternatively, the Mach number discrepancy would mean that the standard DSA from thermal pool electrons fails to explain the origin of a fraction of radio relics and other mechanisms, either the presence of seed electrons or a modification of the standard DSA, are required.
Possible explanations include injection intermittency, where particle acceleration is not continuous but varies over time due to the fact that weak shocks ($\mathcal{M}_{cr}\sim2-3$) may become not efficient in accelerating particles if super-critical conditions are not generated at these shocks \citep[e.g][]{Ha2021}.
In this case, continuous injection may not be established, affecting the overall downstream spectrum. 
Another contributing factor could be Alfvénic drift at the shock, which is expected to modify the spectrum of the accelerated particles \citep[e.g][]{Kang2012}.
While a detailed investigation of these effects is beyond the scope of this work, the Mach number discrepancy highlights the complexity of low-Mach number shocks in galaxy clusters and the need for more refined models of shock acceleration.

\subsection{Surface brightness profile across the northern relic}
\label{sec: SB_profiles} 
\begin{figure*}[h!]
     \includegraphics[width=\textwidth]{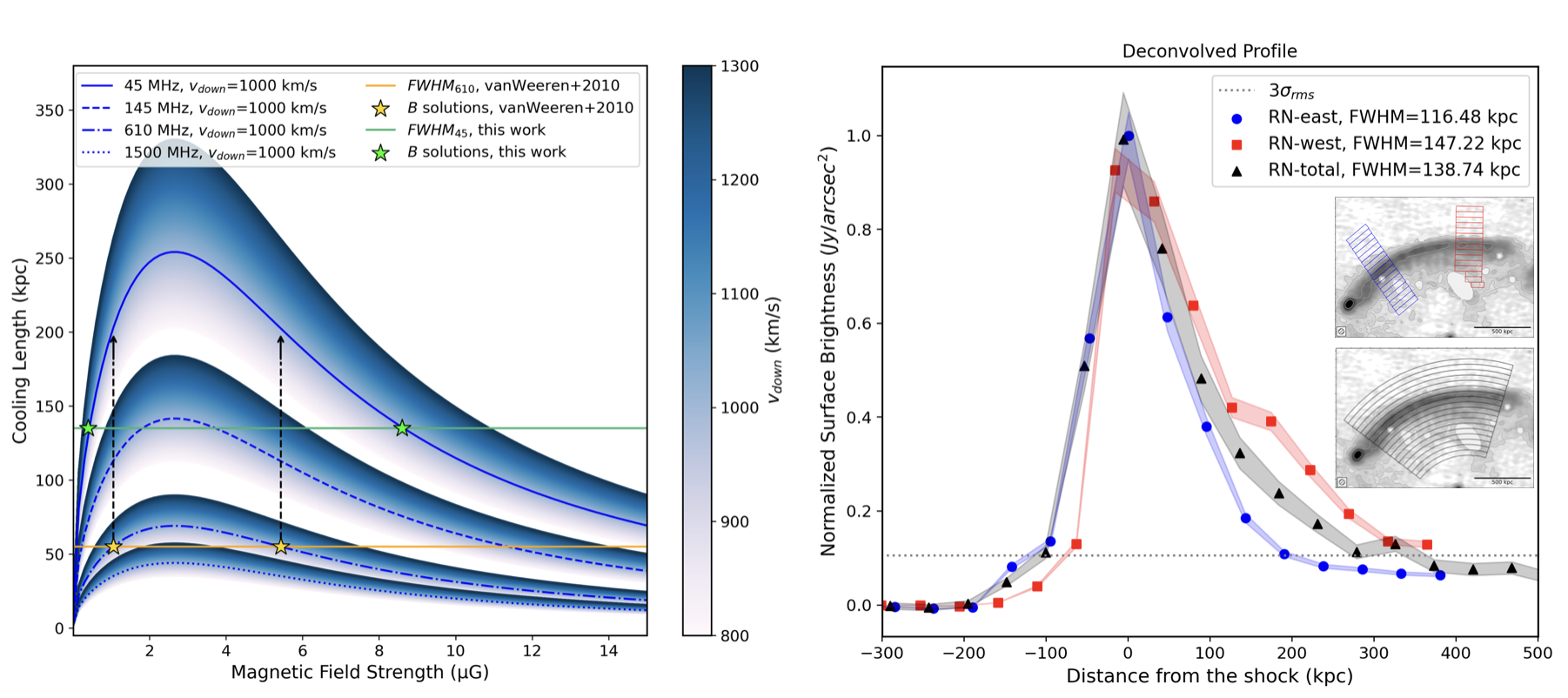}
    \caption{\reviewfirst{Theoretical relic cooling lengths compared with the deconvolved surface brightness profile of RN.
    Left: Relationship between the cooling length (i.e. the relic thickness under ideal conditions of no projection effects and pure radiative losses), magnetic field strength, and varying downstream velocity. Yellow stars represent the B solutions from \cite{vanWeeren2010}, while the green stars are the values from the analysis at 45 MHz. As a comparison, the vertical dashed arrows indicate the expected full width half maximum $\rm FWHM_{\rm  45\,MHz}$ at low frequency, corresponding to the extrapolated values from the relic B solutions derived from $\rm FWHM_{\rm  610\,MHz}$.
    Right: Normalized surface brightness versus distance from the shock. Negative values correspond to the upstream region, while positive ones correspond to the downstream. Data points represent measurements extracted from the LOFAR model-image at 45 MHz along the eastern half (blue), western half (red) and full length (black) of the main relic RN. The dotted horizontal line indicates the image noise level of $\rm 3\sigma_{rms}=1.70\times10^{-5} Jy/arcsec^2$. The insets show the $15\arcsec$-spaced extraction regions, with black 500 kpc scale bar in the bottom right corner.}
    \label{fig:deconv_and_cooling_length}
    }
\end{figure*}

\begin{figure*}
     \includegraphics[width=\textwidth]{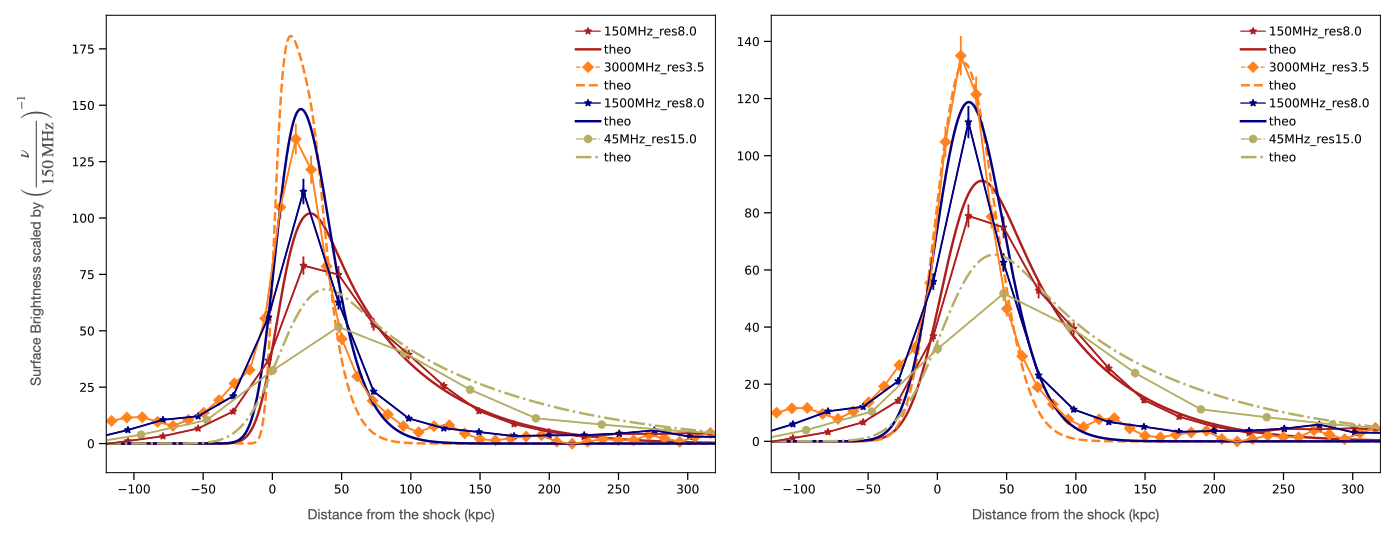}
    \caption{Surface brightness profile across the western half of RN. We applied the model to different frequencies at different resolutions: 3 GHz at $3.5\arcsec$ (orange), 1.5 GHz at $8\arcsec$ (blue), 150 MHz at $8\arcsec$ (red), and 45 MHz at $15\arcsec$.
    Left: Log-normal magnetic field distribution ($\rm B_0=0.3\,\mu G$, $\rm log(\sigma)=1.65$) and $\rm \Psi = 12^{\circ}$ projection.
    Right: Log-normal magnetic field distribution ($\rm B_0=0.3\,\mu G$, $\rm log(\sigma)=1.65$), $\rm \Psi = 12^{\circ}$ projection and additional wiggle-effect with $\sigma=15$ kpc.
    }
    \label{fig:SB_profile_model}
\end{figure*}

Due to its simple morphology and minimal projection effects, the modelling of RN has been extensively studied to infer the physical properties of the shock dynamics. 
Low-frequency observations are particularly sensitive to re-acceleration mechanisms and variations in the magnetic field, making them crucial for understanding the relic physics.
In this section, we analyse the surface brightness (SB) profile of RN, focusing on 45 MHz data, and compare it with results at higher frequencies and theoretical models.

In the simple and idealized scenario of a planar shock observed perfectly edge-on (i.e. without projection effects) and assuming a uniform magnetic field, the width of the relic can be estimated based on spectral ageing principles. 
If no additional acceleration mechanisms are present in the downstream region, the width of the radio shock is entirely determined by spectral ageing due to synchrotron and IC losses at a given frequency $\nu$ as $\tau_{\rm loss} \propto B^{0.5}/ \left[ (B^2+B_{CMB}^2)\times (\nu(1+z))^{0.5}\right]$, where $B$ is the magnetic field strength and $B_{CMB}\propto(1+z)^2$ is the Cosmic Microwave Background equivalent magnetic field strength. 
From dynamical considerations, the relic's cooling length can be derived as $l_{\rm cool}=\tau_{\rm loss}\times \upsilon_{d}$, where the shock's downstream velocity is $\rm \upsilon_d=c_{s}(\mathcal{M}^2+3)/(4\mathcal{M})\propto \sqrt{\gamma T}(\mathcal{M}^2+3)/(4\mathcal{M})$.
The relationship between $l_{\rm cool}$ and $B$ shows that the magnetic field can be estimated from the relic width \citep{Markevitch2005_A520}. Conversely, from independent measurements of the magnetic field, the expected width of the relic can be predicted and compared with observations.
\reviewfirst{Figure~\ref{fig:deconv_and_cooling_length} (left panel)} shows the dependence of the cooling length $l_{\rm cool}$ on the magnetic field for varying $\rm \upsilon_d \in[800,1300]\, km/s $. 
This range keeps into account all possible combinations of the lowest and highest Mach numbers literature values derived from radio observations (\cref{tab:RNRS_Mach_numbers}), combined with pre-shock temperatures from X-ray study: $\sim 2.7$ keV \citep{Akamatsu2015} and $\sim 3.35$ keV \citep{Ogrean2014Chandra}.
This is then calculated at fixed cluster redshift and for four reference frequencies (45, 145, 610, and 1500 MHz, from top to bottom, respectively).
Using the LOFAR LBA $15\arcsec$ image, we extracted surface brightness profiles along the main relic to estimate its width. To measure the intrinsic width of the relic and mitigate the effects of beam convolution, we used the clean component model image. 
This approach allows us to directly obtain the deconvolved SB profiles, leading to a reliable estimate of the relic intrinsic width. 
In \cref{fig:deconv_and_cooling_length} (right panel) we show the deconvolved profiles, extracted in $70\times15\arcsec$ boxes, along the west (red) and east (blue) part of the relic. Additionally, we employed circular annuli to obtain an average SB measure along the full relic size $\rm\sim 1.1\,Mpc$ (black). We normalized the values and shifted them so that the shock position aligns with the zero: negative values correspond to the upstream region, while positive ones correspond to the downstream part. We blank the compact sources that might contaminate the profiles with the help of higher frequency and higher resolution images, where these sources are more clearly visible.
We immediately notice that while the RN-west profile shows a rapid increase followed by a gradual decrease, the RN-east profile appears remarkably symmetrical. 
The asymmetry in the western profile (red) is consistent with expectations for an edge-on relic moving outwards: a sharp rise in radio luminosity at the shock location, followed by a frequency-dependent decline that becomes steeper at higher frequencies due to radiative cooling $t\propto1/E$.
On the other hand, the symmetry and relatively smooth profile of the eastern side, with wings extending on either side of the peak, is a strong deviation from this expected model. 
This duality of the east- and west- sides corresponds to the morphological difference discussed in \cref{sec:mach_number} and \cref{sec:radio_morphology}.
The overall profile (black) maintains the broader upstream profile while representing an average of the east (blue) and west (red) curves in the downstream area.
While we have masked sources that could affect the profile, it is important to note that source B is very extended and likely interacts with the downstream emission on the west side. Therefore, the red curve might be influenced by this interaction, meaning it does not purely represent the relic downstream emission.
The deconvolved overall profile has an FWHM of 138 kpc at 45 MHz. This width would imply magnetic field values of $0.4 \,\mu G$ or $8.5\,\mu G$, as each curve of \cref{fig:deconv_and_cooling_length} has two solutions for a given width value.
If projection effects play a significant role, the intrinsic profile would be narrower, implying magnetic field values of $B<0.4\,\mu G$ or $B>8.5\,\mu G$. 
Based on constraints from IC emission, equipartition arguments and Faraday rotation measurements \citep{Stroe2014_Spectral_age_modelling, vanWeeren2010}, we can exclude the lower solution. 
Despite being based on ideal assumptions, these theoretical estimates are useful to constrain the possible range of magnetic field strengths at the relic location, at approximately 1.5 Mpc away from the cluster centre. 
Only a few other studies provide magnetic field estimates for this galaxy cluster \citep{vanWeeren2010, Donnert2016,DiGennaro2021}.
Similar considerations were put forward in \cite{vanWeeren2010} who found a deconvolved FWHM of 55 kpc at 610 MHz, leading to $B\lesssim 1.2\,\mu G$ or $\rm B\gtrsim 5\,\mu G$.
Extrapolating these magnetic field values to lower frequencies, such as 45 MHz, we would expect the FWHM to broaden significantly, resulting in width estimates of approximately $ l_{\rm cool,45}\lesssim \rm 200\,kpc \gg  FMHW_{deconv,45} \sim 138\,kpc$.
It is important to note that the FWHM is only an indicator of the cooling length. In the case of the 610 MHz data, the relic width remains relatively constant across the entire profile. Thus, the FWHM is a reliable estimator. Conversely, at 45 MHz, we observe a more pronounced broadening downstream (see \cref{app:profile_shape_analysis} for an appropriate discussion about the shape of the profile). 
If we consider only the downstream extension (from the shock front to the location where the intensity drops to 10\% of its peak), we obtain a width of approximately 200 kpc, which is more in line with the extrapolation from \cite{vanWeeren2010}.
The 10\% intensity level also roughly corresponds to the $\rm 3\sigma_{rms}$ threshold, which defines the limit for a reliable measurement. If we consider this and the maximum downstream extent, it provides a lower limit on B, confirming that $\rm B>5 \mu G$ is in agreement with \cite{vanWeeren2010}.
Another independent measure of the magnetic field strength was obtained by \cite{DiGennaro2021}, who used Faraday depolarization to estimate the turbulent magnetic field component $\rm B_{turb}\sim6\,\mu G$.
Given the assumptions and uncertainties related to these measurements, we can conclude that our estimates are broadly consistent with previous studies. We can constrain the magnetic field at the relic in the range of 5 to 10 $\mu G$.

Despite the well-defined morphology that is characteristic of the northern relic, RN, in \CIZA, the shock geometry in a merging system is inherently more complex. 
For example, higher-resolution and frequency observations reveal a fragmented and irregular shock surface on smaller scales \citep[see Fig. 7 in][]{DiGennaro2018}, which suggests a more complex shock structure. 
Additionally, we observe a broadening in the upstream region of the profile (see \cref{fig:deconv_and_cooling_length}, right). This raises the question of whether this is the result of projection effects or an intrinsic shock property.
For these reasons, we devised a model that incorporates projection effects, possible downstream magnetic field variation and shock surface irregularities.

We model the surface brightness profile of the main relic RN, using a framework that considers, both, the physics of particle acceleration and the effects of geometry and magnetic field variations. 
We assume the shock front is a spherically shaped cap with a uniform Mach number.
Each surface element of this spherical cap acts as the origin of a downstream emission profile,
computed based on an injection spectrum for the relativistic electron population and downstream conditions parametrized by the Mach number and post-shock gas temperature $(k_{B}T)$.
The cosmic-ray electrons follow a power-law momentum distribution at the shock front, with their synchrotron emission decreasing with downstream distance due to radiative losses.
With minor modifications, we follow the formalism of \cite{HoeftBruggen2007}.
Projection effects are accounted for by integrating along the line of sight for a given curvature radius ($\rm R=1.5\,kpc$) and maximum opening angle $\Psi$ of the spherical cap, resulting in a realistic surface brightness distribution.
We extend the model by \cite{DiGennaro2018} by incorporating additional data at 45 MHz, the lowest frequency available, which enables us to better constrain the model parameters and test its applicability in the extreme low-frequency regime.

It has been shown \citep[e.g.][]{Hoeft2011, Lee2024} that shock surfaces in galaxy clusters are not smooth, but exhibit small-scale irregularities. This is also observed in other clusters \citep[e.g.][]{deGasperin2022, Rajpurohit2020_Toothbrush}.
The deviations from a smooth shock surface are poorly constrained. To mimic them as part of our model, we assume that the radius in the spherical cap model has a Gaussian scatter. The width of this scatter, which we will refer to as `wiggles', becomes an additional free parameter in our model. In contrast the wiggles cause a symmetric broadening in the projected shock surface profile.  
We extracted surface brightness profiles in $70\arcsec\times15\arcsec$ boxes along the eastern half of the RN, using 45, 145, 1550 and 3000 MHz images. 
We exploit the multi-frequency dataset to take advantage of different achievable resolutions.

We start by considering the effects of projection alone. Primarily, we relied on the highest $3\arcsec$-resolution image at 3 GHz to constrain the maximum possible projection allowed. 
At lower resolutions, the broader and smoother surface brightness profiles would allow the opening angle $\Psi$ to vary more freely in the parameter space.
The opening angle that best fits the data is $\Psi= 12^{\circ} $, which implies an injection up to about 33 kpc (projected size) in the downstream direction,
keeping the magnetic field constant at $\rm B=3\,\mu G$.
We demonstrate that even with a slightly different opening angle  $\Psi= 18^{\circ} $, the profile would strongly deviate from observations, creating a prominent downstream tail (\cref{fig:SB_profile_model_18deg}). 
After fixing the opening angle, we applied the model to different frequencies at different resolutions to assess its consistency across the dataset. In \cref{fig:SB_profile_model}, we compare 3 GHz at $3.5\arcsec$ (orange), 1.5 GHz at $8\arcsec$ (blue), 150 MHz at $8\arcsec$ (red) and 45 MHz at $15\arcsec$.
In order to test the standard DSA scenario, we fix the Mach number to $\mathcal{M}=4.8$, as derived from the $\rm \alpha_{int}$ in this work (\cref{sec:mach_number}).

Moreover, in such a complex environment as the one of the clusters undergoing mergers, it is reasonable to believe that the magnetic fields do not remain uniform across Mpc-scale shock fronts. 
To capture the turbulent nature of the ICM, the magnetic field is modelled as a log-normal distribution, introducing variability in the downstream emission profiles originating from each surface element.
We find that the best-fit model has a mean magnetic field strength of $\rm B_0 = 0.3\, \mu G$ and scatter $\rm log(\sigma)=1.65$. (\cref{fig:SB_profile_model}, left). This implies that the magnetic field varies in a range from approximately $3\times 10^{-3}$ to $\rm 16\, \mu G$. 
This model is able to match the observed profiles in the downstream area reasonably well, while it substantially fails to reproduce the part of the profile ahead of the peak, which shows excess respect for all the modeled profiles \cref{fig:SB_profile_model}.
For this reason, we added wiggles with a Gaussian broadening to the original spherical cap model to test its impact on the fit. 
The results are shown in \cref{fig:SB_profile_model} (right), where we applied a Gaussian broadening characterized by $\rm \sigma = 15\,kpc$ to match the width of the 3.5\, arcsec profile. This helps in smoothing the models in the upstream region, however a significant discrepancy is still present. 
Notably, all profiles exhibit a systematic excess in the upstream region extending for approximately 100 kpc, consistent with the average profile shown in \cref{fig:deconv_and_cooling_length} (right, black line).
Overall, our modelling efforts successfully reproduce the general shape of the observed surface brightness profiles, particularly in the downstream region. However, some discrepancies remain, particularly in the upstream region, where all modelled profiles show a systematic excess compared to observations.

\subsection{R5}
\label{subsec:R5}
\begin{figure}
    \includegraphics[width=0.85\columnwidth]{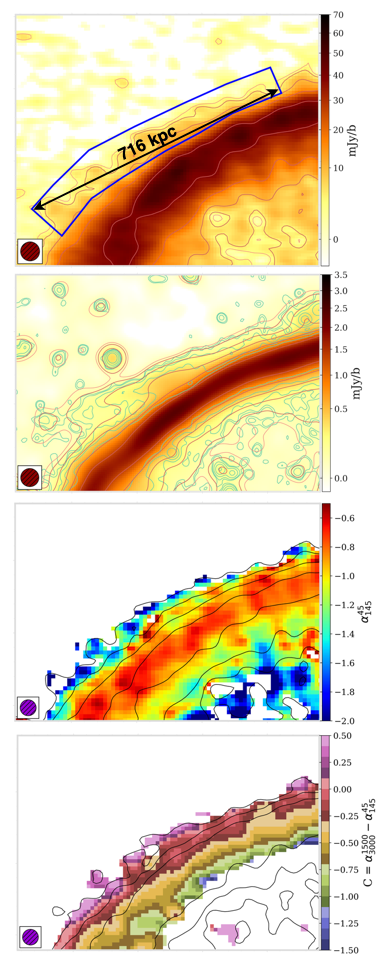}
    \caption{Zoom-in on the R5 emission regio. From top to bottom: 45 MHz image at $15\arcsec$ resolution; 1.5 GHz image convolved to the same $15\arcsec$ resolution and green contours indicating the $5\arcsec$ higher resolution data; spectral index map between 45 and 145 MHz from \cref{fig:spectral_index_map}, 4-frequency (45, 145, 1500, 3000 MHz) curvature map from \cref{fig:curvature_map}}
    \label{fig:R5_zoomin}
\end{figure}

As mentioned in \cref{sec:radio_morphology}, in the north-west part of the northern relic, we find an excess of emission that extends for $\rm 716\,kpc$ and has an average width of $\rm \sim 100\,kpc$ (see \cref{fig:R5_zoomin}). 
R5 was first identified at GHz frequencies by \cite{DiGennaro2018} and recently confirmed at 400 and 675 MHz by \cite{Raja2024}. 
In our LOFAR LBA images, R5 appears irregular and less confined compared to the distinct spike structure seen at higher frequencies. 
A similar diffuse appearance is observed in LOFAR HBA images, suggesting intrinsic differences in the emission at low frequencies rather than being solely a consequence of resolution limitations, although resolution likely contributes to the smoothing of finer substructures.
This is further demonstrated in \cref{fig:R5_zoomin} (second panel), where a comparison between the 15\arcsec-beam smoothed image (with corresponding  red contours) and the $5\arcsec$ contours (green) at higher frequencies shows that the spiked feature remains visible even after smoothing.
Moreover, we note that the width of R5 corresponds to $\sim 2.5$ times the beam size (that is, $15\arcsec$ nominal resolution at 45 MHz), thus we exclude the possibility of it purely being the result of the low resolution of the telescope.
This area has steep spectral spectral index values $\alpha_{45}^{145}\in[-1.2,-0.9]$ range (see spectral index map in \cref{fig:R5_zoomin}). Even if the relative errors are higher at the edges of the emission, with $\rm \Delta \alpha_{45}^{145}\sim 0.2-0.3$ there is a clear distinction between R5 area and the main shock front area. This is also confirmed by the local spectral index analysis along the RN outer edge of \cref{sec:mach_number} (\cref{fig:spectral_index_injection}).
We calculated the integrated spectral index within the R5 region (blue box of \cref{fig:R5_zoomin}, top panel) by fitting a single power-law to the data and obtaining a best-fit value of $\alpha_{int}^{R5}=-0.90\pm0.03$.  Due to the extremely low surface brightness of R5 we used only the 45, 145, 1500 and 3000 MHz images, where a clear detection above $\rm 3\sigma_{rms}$ threshold was visible.
The resulting power-law fit is consistent with the two-frequency spectral index reported by \cite{DiGennaro2018} and aligns with the curvature map (\cref{fig:R5_zoomin}, bottom panel), which shows $C=0$ in the R5 region, supporting the absence of significant spectral curvature.
Hints of polarized emission have been detected for R5, with a degree of polarization of $\sim35\%$ (at 3 GHz) and $\sim30\%$ at 1.5 GHz \citep{DiGennaro2021}. This level of polarization is consistent with what is typically expected for radio relics, supporting the idea that R5 is associated with a shock-related structure. However, its polarization fraction is lower than that of the main relic, suggesting potential differences in shock strength.
The nature of R5 remains unclear, with two primary interpretations: either it is separate relic seen in projection, or it represents broader upstream substructure of the main northern relic, extending the shock surface in that direction.

\section{Conclusions}
\label{sec:conclusion}
We presented the first ultra-low frequency observations of the Sausage relic at 45 MHz using the LOFAR telescope, achieving a thermal-noise limited image with a noise level of $ \rm 1.5 \, mJy/beam $ at $ 15\arcsec $ resolution. These observations mark the lowest radio frequency at which this cluster has been studied to date and demonstrate the capabilities of the LOFAR LBA calibration pipeline for producing high-fidelity images of diffuse radio sources.
\begin{enumerate}
   \item The 45 MHz observations reveal a complex system of diffuse, relic-like sources, extending beyond the well-known northern and southern relics. The northern relic shows a characteristic arc-like morphology with a projected linear size of $\rm \sim 2.2 \, Mpc \times 760 \, kpc$, while the southern relic exhibits a more irregular shape for a total extent of $\rm \sim 1.5 \, Mpc \times 520 \, kpc$. 
   The morphology of the relics not only appears larger but reveals connections between substructures that would appear disconnected at higher frequencies. Most sources (R1, R2, RS1, RS2, RS3, RS4), appear more fragmented with increasing frequency.
   \item Spectral index maps between 45 MHz and 144 MHz reveal a clear spectral gradient in both relics, where the spectral index steepens from the outer edge towards the cluster centre. RN shows a gradient from $\alpha \sim -0.7$ to $\sim -1.8$, while RS exhibits a broader variation, with the southeastern region (RS-east) showing the expected steepening (from $\alpha \sim -0.5$ to $\sim -2$) while the north-western part (RS-west) remaining consistently steep ($\sim -1.8$) even at the outermost edge, with no inward steepening. These broad steep-spectrum north-western areas of RS explain why RS-west fades quickly at higher frequencies ($\rm \nu\gtrsim600\,MHz$), where primarily the southern edge remains bright and highlights the importance of low frequencies in tracing these faint parts of relics.
   Additionally, the spectral index maps provide evidence that R1 consists of two distinct regions: (i) R1-east with a flatter spectrum ($\alpha \sim -0.8$) and (ii) R1-west, which is aligned with R4 ($\alpha \sim -1.3$). The patchy morphology and mixed spectral indices of R2 indicate that this source may be seen at a larger inclination.
   \item The integrated spectral index of the northern relic is $\alpha_\mathrm{int}^N = -1.09 \pm 0.03$, while the southern relic has a steeper integrated index of $\alpha_\mathrm{int}^S = -1.34 \pm 0.03$. Using these indices, Mach numbers were derived as $\mathcal{M}_N = 4.8 \pm 0.8$ for the northern relic and $\mathcal{M}_S = 2.6 \pm 0.1$ for the southern relic.
   \item Access to ultra-low radio frequencies, which are less affected by radiative losses, allows us to measure the injection spectral index along the relic edges. By applying statistical methods to identify distinct spectral index populations, we determined injection indices of $\bar{\alpha}_{\mathrm{inj}}^N  = -0.76 \pm 0.08$ and $\bar{\alpha}_{\mathrm{inj}}^S = -0.77 \pm 0.16$, corresponding to Mach numbers of $\mathcal{M}_N = 2.9 \pm 0.4$ and $\mathcal{M}_S = 2.9 \pm 0.8$. The close agreement of these Mach numbers suggests that the shocks causing both relics have comparable strengths, despite their differing morphologies.
   \item Using multi-frequency observations, we produced a four-frequency spectral curvature map from 45 MHz to 3 GHz. The curvature analysis of the northern relic reveals a perfectly flat spectrum with $C \sim 0$ at the outer edge, and a gradient up to $C \sim -1.5$ in the downstream region, in agreement with expectations from DSA.
   The southern relic instead shows a smaller gradient with values $C \sim -0.2$ to $-0.5$, suggesting it might be influenced by additional effects, such as turbulence or projection effects.
   \item Finally, we used a toy model to simulate the surface brightness profile of RN to gain insights into cluster dynamics, projection effects, and the magnetic field strength at the shock location. 
   Our modelling indicates that RN is best represented by a scenario with minimal projection effects (low opening angle $\Psi=12^{\circ}$), spatial variations in the downstream magnetic field, modelled as a log-normal distribution with characteristic strength of $\rm B_0=0.3\,\mu G$ and scatter $\log(\sigma)=1.65,$ and Gaussian broadening of the shock surface with $\rm \sigma=15\,kpc$. Even if the additional broadening helps in smoothing the upstream profile, some discrepancies remain, especially in the upstream region, where an excess of emission systematically persists at all frequencies.
   Moreover, we stress that the toy model proposed in this section serves to explore the fundamental dependencies of the surface brightness profile on the geometry of the system, projection, and magnetic field variations. A fully accurate theoretical representation would require dedicated tailored simulations, which are beyond the scope of this work.

\end{enumerate}

\begin{acknowledgements}
\reviewfirst{We thank the anonymous referee for a very constructive report.}
MB acknowledges funding by the Deutsche Forschungsgemeinschaft (DFG, German Research Foundation) under Germany's Excellence Strategy -- EXC 2121 ``Quantum Universe'' -- 390833306.
FdG and GDG acknowledges support from the ERC Consolidator Grant ULU 101086378.
\reviewfirst{
The new data presented in this paper are obtained with the LOw Frequency ARray (LOFAR).
LOFAR data products were provided by the LOFAR Surveys Key Science project (LSKSP; \url{https://lofar-surveys.org/}) and were derived from observations with the International LOFAR Telescope (ILT). LOFAR \citep{vanHaarlem2013} is the Low Frequency Array designed and constructed by ASTRON. It has observing, data processing, and data storage facilities in several countries, which are owned by various parties (each with their own funding sources), and which are collectively operated by the ILT foundation under a joint scientific policy. The ILT resources have benefited from the following recent major funding sources: CNRS-INSU, Observatoire de Paris and Université d’Orléans, France; BMBF, MIWF-NRW, MPG, Germany; Science Foundation Ireland (SFI), Department of Business, Enterprise and Innovation (DBEI), Ireland; NWO, The Netherlands; The Science and Technology Facilities Council, UK; Ministry of Science and Higher Education, Poland; The Istituto Nazionale di Astrofisica (INAF), Italy.
This research made use of the Dutch national e-infrastructure with support of the SURF Cooperative (e-infra 180169) and the LOFAR e-infra group. The Jülich LOFAR Long Term Archive and the German LOFAR network are both coordinated and operated by the Jülich Supercomputing Centre (JSC), and computing resources on the supercomputer JUWELS at JSC were provided by the Gauss Centre for Supercomputing e.V. (grant CHTB00) through the John von Neumann Institute for Computing (NIC).
This research made use of the University of Hertfordshire high-performance computing facility and the LOFAR-UK computing facility located at the University of Hertfordshire and supported by STFC [ST/P000096/1]. This research made use of the LOFAR-IT computing infrastructure supported and operated by INAF, including the resources within the PLEIADI special “LOFAR” project by USC-C of INAF, and by the Physics Dept. of Turin University (under the agreement with Consorzio Interuniversitario per la Fisica Spaziale) at the C3S Supercomputing Centre, Italy.\\
This paper also made use of data obtained with the Karl G. Jansky Very Large Array (VLA), the Giant Metrewave Radio Telescope (GMRT) and Westerbork Synthesis Radio Telescope (WSRT). 
The National Radio Astronomy Observatory is a facility of the National Science Foundation operated under cooperative agreement by Associated Universities, Inc.
We thank the staff of the GMRT that made these observations possible. GMRT is run by the National Centre for Radio Astrophysics of the Tata Institute of Fundamental Research.
The WSRT is operated by ASTRON (Netherlands Institute for Radio Astronomy) with support from the Netherlands Foundation for Scientific Research NWO.
}
\end{acknowledgements}

\bibliographystyle{aa} 
\bibliography{mybibl} 

\begin{thebibliography}{79}
\expandafter\ifx\csname natexlab\endcsname\relax\def\natexlab#1{#1}\fi

\bibitem[{{Akaike}(1974)}]{Akaike1974}
{Akaike}, H. 1974, IEEE Transactions on Automatic Control, 19, 716

\bibitem[{{Akamatsu} \& {Kawahara}(2013)}]{Akamatsu2013}
{Akamatsu}, H. \& {Kawahara}, H. 2013, \pasj, 65, 16

\bibitem[{{Akamatsu} {et~al.}(2015){Akamatsu}, {van Weeren}, {Ogrean},
  {Kawahara}, {Stroe}, {Sobral}, {Hoeft}, {R{\"o}ttgering}, {Br{\"u}ggen}, \&
  {Kaastra}}]{Akamatsu2015}
{Akamatsu}, H., {van Weeren}, R.~J., {Ogrean}, G.~A., {et~al.} 2015, \aap, 582,
  A87

\bibitem[{{Blandford} \& {Eichler}(1987)}]{Blandford1987}
{Blandford}, R. \& {Eichler}, D. 1987, \physrep, 154, 1

\bibitem[{{Blandford} \& {Ostriker}(1978)}]{BlandfordOstriker1978}
{Blandford}, R.~D. \& {Ostriker}, J.~P. 1978, \apjl, 221, L29

\bibitem[{{Bonafede} {et~al.}(2009){Bonafede}, {Giovannini}, {Feretti},
  {Govoni}, \& {Murgia}}]{Bonafede2009}
{Bonafede}, A., {Giovannini}, G., {Feretti}, L., {Govoni}, F., \& {Murgia}, M.
  2009, \aap, 494, 429

\bibitem[{{Botteon} {et~al.}(2020){Botteon}, {Brunetti}, {Ryu}, \&
  {Roh}}]{Botteon2020_shock_acceleration}
{Botteon}, A., {Brunetti}, G., {Ryu}, D., \& {Roh}, S. 2020, \aap, 634, A64

\bibitem[{{Brunetti} \& {Jones}(2014)}]{Brunetti&Jones2014}
{Brunetti}, G. \& {Jones}, T.~W. 2014, International Journal of Modern Physics
  D, 23, 1430007

\bibitem[{{Condon} {et~al.}(1998){Condon}, {Cotton}, {Greisen}, {Yin},
  {Perley}, {Taylor}, \& {Broderick}}]{Condon1998}
{Condon}, J.~J., {Cotton}, W.~D., {Greisen}, E.~W., {et~al.} 1998, \aj, 115,
  1693

\bibitem[{{Dawson} {et~al.}(2015){Dawson}, {Jee}, {Stroe}, {Ng}, {Golovich},
  {Wittman}, {Sobral}, {Br{\"u}ggen}, {R{\"o}ttgering}, \& {van
  Weeren}}]{Dawson2015}
{Dawson}, W.~A., {Jee}, M.~J., {Stroe}, A., {et~al.} 2015, \apj, 805, 143

\bibitem[{{de Gasperin} {et~al.}(2020{\natexlab{a}}){de Gasperin}, {Brunetti},
  {Br{\"u}ggen}, {van Weeren}, {Williams}, {Botteon}, {Cuciti}, {Dijkema},
  {Edler}, {Iacobelli}, {Kang}, {Offringa}, {Orr{\'u}}, {Pizzo}, {Rafferty},
  {R{\"o}ttgering}, \& {Shimwell}}]{deGasperin2020}
{de Gasperin}, F., {Brunetti}, G., {Br{\"u}ggen}, M., {et~al.}
  2020{\natexlab{a}}, \aap, 642, A85

\bibitem[{{de Gasperin} {et~al.}(2019){de Gasperin}, {Dijkema}, {Drabent},
  {Mevius}, {Rafferty}, {van Weeren}, {Br{\"u}ggen}, {Callingham}, {Emig},
  {Heald}, {Intema}, {Morabito}, {Offringa}, {Oonk}, {Orr{\`u}},
  {R{\"o}ttgering}, {Sabater}, {Shimwell}, {Shulevski}, \&
  {Williams}}]{deGasperin2019}
{de Gasperin}, F., {Dijkema}, T.~J., {Drabent}, A., {et~al.} 2019, \aap, 622,
  A5

\bibitem[{{de Gasperin} {et~al.}(2023){de Gasperin}, {Edler}, {Williams},
  {Callingham}, {Asabere}, {Br{\"u}ggen}, {Brunetti}, {Dijkema}, {Hardcastle},
  {Iacobelli}, {Offringa}, {Norden}, {R{\"o}ttgering}, {Shimwell}, {van
  Weeren}, {Tasse}, {Bomans}, {Bonafede}, {Botteon}, {Cassano}, {Chy{\.z}y},
  {Cuciti}, {Emig}, {Kadler}, {Miley}, {Mingo}, {Oei}, {Prandoni}, {Schwarz},
  \& {Zarka}}]{deGasperin2023}
{de Gasperin}, F., {Edler}, H.~W., {Williams}, W.~L., {et~al.} 2023, \aap, 673,
  A165

\bibitem[{{de Gasperin} {et~al.}(2015){de Gasperin}, {Intema}, {van Weeren},
  {Dawson}, {Golovich}, {Wittman}, {Bonafede}, \&
  {Br{\"u}ggen}}]{deGasperin2015}
{de Gasperin}, F., {Intema}, H.~T., {van Weeren}, R.~J., {et~al.} 2015, \mnras,
  453, 3483

\bibitem[{{de Gasperin} {et~al.}(2022){de Gasperin}, {Rudnick}, {Finoguenov},
  {Wittor}, {Akamatsu}, {Br{\"u}ggen}, {Chibueze}, {Clarke}, {Cotton},
  {Cuciti}, {Dom{\'\i}nguez-Fern{\'a}ndez}, {Knowles}, {O'Sullivan}, \&
  {Sebokolodi}}]{deGasperin2022}
{de Gasperin}, F., {Rudnick}, L., {Finoguenov}, A., {et~al.} 2022, \aap, 659,
  A146

\bibitem[{{de Gasperin} {et~al.}(2020{\natexlab{b}}){de Gasperin}, {Vink},
  {McKean}, {Asgekar}, {Avruch}, {Bentum}, {Blaauw}, {Bonafede}, {Broderick},
  {Br{\"u}ggen}, {Breitling}, {Brouw}, {Butcher}, {Ciardi}, {Cuciti}, {de Vos},
  {Duscha}, {Eisl{\"o}ffel}, {Engels}, {Fallows}, {Franzen}, {Garrett},
  {Gunst}, {H{\"o}randel}, {Heald}, {Hoeft}, {Iacobelli}, {Koopmans},
  {Krankowski}, {Maat}, {Mann}, {Mevius}, {Miley}, {Morganti}, {Nelles},
  {Norden}, {Offringa}, {Orr{\'u}}, {Paas}, {Pandey}, {Pandey-Pommier},
  {Pekal}, {Pizzo}, {Reich}, {Rowlinson}, {Rottgering}, {Schwarz}, {Shulevski},
  {Smirnov}, {Sobey}, {Soida}, {Steinmetz}, {Tagger}, {Toribio}, {van Ardenne},
  {van der Horst}, {van Haarlem}, {van Weeren}, {Vocks}, {Wucknitz}, {Zarka},
  \& {Zucca}}]{deGasperin2020_ATeam}
{de Gasperin}, F., {Vink}, J., {McKean}, J.~P., {et~al.} 2020{\natexlab{b}},
  \aap, 635, A150

\bibitem[{{de Gasperin} {et~al.}(2021){de Gasperin}, {Williams}, {Best},
  {Br{\"u}ggen}, {Brunetti}, {Cuciti}, {Dijkema}, {Hardcastle}, {Norden},
  {Offringa}, {Shimwell}, {van Weeren}, {Bomans}, {Bonafede}, {Botteon},
  {Callingham}, {Cassano}, {Chy{\.z}y}, {Emig}, {Edler}, {Haverkorn}, {Heald},
  {Heesen}, {Iacobelli}, {Intema}, {Kadler}, {Ma{\l}ek}, {Mevius}, {Miley},
  {Mingo}, {Morabito}, {Sabater}, {Morganti}, {Orr{\'u}}, {Pizzo}, {Prandoni},
  {Shulevski}, {Tasse}, {Vaccari}, {Zarka}, \&
  {R{\"o}ttgering}}]{deGasperin2021}
{de Gasperin}, F., {Williams}, W.~L., {Best}, P., {et~al.} 2021, \aap, 648,
  A104

\bibitem[{{Di Gennaro} {et~al.}(2018){Di Gennaro}, {van Weeren}, {Hoeft},
  {Kang}, {Ryu}, {Rudnick}, {Forman}, {R{\"o}ttgering}, {Br{\"u}ggen},
  {Dawson}, {Golovich}, {Hoang}, {Intema}, {Jones}, {Kraft}, {Shimwell}, \&
  {Stroe}}]{DiGennaro2018}
{Di Gennaro}, G., {van Weeren}, R.~J., {Hoeft}, M., {et~al.} 2018, \apj, 865,
  24

\bibitem[{{Di Gennaro} {et~al.}(2021){Di Gennaro}, {van Weeren}, {Rudnick},
  {Hoeft}, {Br{\"u}ggen}, {Ryu}, {R{\"o}ttgering}, {Forman}, {Stroe},
  {Shimwell}, {Kraft}, {Jones}, \& {Hoang}}]{DiGennaro2021}
{Di Gennaro}, G., {van Weeren}, R.~J., {Rudnick}, L., {et~al.} 2021, \apj, 911,
  3

\bibitem[{{Dominguez-Fernandez} {et~al.}(2021){Dominguez-Fernandez}, {Bruggen},
  {Vazza}, {Banda-Barragan}, {Rajpurohit}, {Mignone}, {Mukherjee}, \&
  {Vaidya}}]{Dominguez-Fernandez2021}
{Dominguez-Fernandez}, P., {Bruggen}, M., {Vazza}, F., {et~al.} 2021, \mnras,
  500, 795

\bibitem[{{Dom{\'\i}nguez-Fern{\'a}ndez}
  {et~al.}(2024{\natexlab{a}}){Dom{\'\i}nguez-Fern{\'a}ndez}, {Ryu}, \&
  {Kang}}]{Dominguez-Fernandez2024}
{Dom{\'\i}nguez-Fern{\'a}ndez}, P., {Ryu}, D., \& {Kang}, H.
  2024{\natexlab{a}}, \aap, 685, A68

\bibitem[{{Dom{\'\i}nguez-Fern{\'a}ndez}
  {et~al.}(2024{\natexlab{b}}){Dom{\'\i}nguez-Fern{\'a}ndez}, {Ryu}, \&
  {Kang}}]{Dom-Fern2024}
{Dom{\'\i}nguez-Fern{\'a}ndez}, P., {Ryu}, D., \& {Kang}, H.
  2024{\natexlab{b}}, \aap, 685, A68

\bibitem[{{Donnert} {et~al.}(2016){Donnert}, {Stroe}, {Brunetti}, {Hoang}, \&
  {Roettgering}}]{Donnert2016}
{Donnert}, J.~M.~F., {Stroe}, A., {Brunetti}, G., {Hoang}, D., \&
  {Roettgering}, H. 2016, \mnras, 462, 2014

\bibitem[{{Drury}(1983)}]{Drury1983}
{Drury}, L.~O. 1983, Reports on Progress in Physics, 46, 973

\bibitem[{{Edler} {et~al.}(2022){Edler}, {de Gasperin}, {Brunetti}, {Botteon},
  {Cuciti}, {van Weeren}, {Cassano}, {Shimwell}, {Br{\"u}ggen}, \&
  {Drabent}}]{Edler2022}
{Edler}, H.~W., {de Gasperin}, F., {Brunetti}, G., {et~al.} 2022, \aap, 666, A3

\bibitem[{{Ensslin} {et~al.}(1998){Ensslin}, {Biermann}, {Klein}, \&
  {Kohle}}]{Ensslin1998}
{Ensslin}, T.~A., {Biermann}, P.~L., {Klein}, U., \& {Kohle}, S. 1998, \aap,
  332, 395

\bibitem[{{Ha} {et~al.}(2021){Ha}, {Kim}, {Ryu}, \& {Kang}}]{Ha2021}
{Ha}, J.-H., {Kim}, S., {Ryu}, D., \& {Kang}, H. 2021, \apj, 915, 18

\bibitem[{{Hoang} {et~al.}(2017){Hoang}, {Shimwell}, {Stroe}, {Akamatsu},
  {Brunetti}, {Donnert}, {Intema}, {Mulcahy}, {R{\"o}ttgering}, {van Weeren},
  {Bonafede}, {Br{\"u}ggen}, {Cassano}, {Chy{\.z}y}, {En{\ss}lin}, {Ferrari},
  {de Gasperin}, {Gu}, {Hoeft}, {Miley}, {Orr{\'u}}, {Pizzo}, \&
  {White}}]{Hoang2017}
{Hoang}, D.~N., {Shimwell}, T.~W., {Stroe}, A., {et~al.} 2017, \mnras, 471,
  1107

\bibitem[{{Hoang} {et~al.}(2018){Hoang}, {Shimwell}, {van Weeren}, {Intema},
  {R{\"o}ttgering}, {Andrade-Santos}, {Akamatsu}, {Bonafede}, {Brunetti},
  {Dawson}, {Golovich}, {Best}, {Botteon}, {Br{\"u}ggen}, {Cassano}, {de
  Gasperin}, {Hoeft}, {Stroe}, \& {White}}]{Hoang2018}
{Hoang}, D.~N., {Shimwell}, T.~W., {van Weeren}, R.~J., {et~al.} 2018, \mnras,
  478, 2218

\bibitem[{{Hoeft} \& {Br{\"u}ggen}(2007)}]{HoeftBruggen2007}
{Hoeft}, M. \& {Br{\"u}ggen}, M. 2007, \mnras, 375, 77

\bibitem[{{Hoeft} {et~al.}(2011){Hoeft}, {Nuza}, {Gottl{\"o}ber}, {van Weeren},
  {R{\"o}ttgering}, \& {Br{\"u}ggen}}]{Hoeft2011}
{Hoeft}, M., {Nuza}, S.~E., {Gottl{\"o}ber}, S., {et~al.} 2011, Journal of
  Astrophysics and Astronomy, 32, 509

\bibitem[{{Inchingolo} {et~al.}(2022){Inchingolo}, {Wittor}, {Rajpurohit}, \&
  {Vazza}}]{Inchingolo2022}
{Inchingolo}, G., {Wittor}, D., {Rajpurohit}, K., \& {Vazza}, F. 2022, \mnras,
  509, 1160

\bibitem[{{Intema} {et~al.}(2017){Intema}, {Jagannathan}, {Mooley}, \&
  {Frail}}]{Intema2017}
{Intema}, H.~T., {Jagannathan}, P., {Mooley}, K.~P., \& {Frail}, D.~A. 2017,
  \aap, 598, A78

\bibitem[{{Jee} {et~al.}(2015){Jee}, {Stroe}, {Dawson}, {Wittman}, {Hoekstra},
  {Br{\"u}ggen}, {R{\"o}ttgering}, {Sobral}, \& {van Weeren}}]{Jee2015}
{Jee}, M.~J., {Stroe}, A., {Dawson}, W., {et~al.} 2015, \apj, 802, 46

\bibitem[{{Kang}(2012)}]{Kang2012}
{Kang}, H. 2012, Journal of Korean Astronomical Society, 45, 127

\bibitem[{{Kang}(2016)}]{Kang2016}
{Kang}, H. 2016, Journal of Korean Astronomical Society, 49, 145

\bibitem[{{Kang} \& {Ryu}(2011)}]{Kang_and_Ryu2011}
{Kang}, H. \& {Ryu}, D. 2011, \apj, 734, 18

\bibitem[{{Kardashev}(1962)}]{Kardashev1962}
{Kardashev}, N.~S. 1962, \sovast, 6, 317

\bibitem[{{Kierdorf} {et~al.}(2017){Kierdorf}, {Beck}, {Hoeft}, {Klein}, {van
  Weeren}, {Forman}, \& {Jones}}]{Kierdorf2017}
{Kierdorf}, M., {Beck}, R., {Hoeft}, M., {et~al.} 2017, \aap, 600, A18

\bibitem[{{Kocevski} {et~al.}(2007){Kocevski}, {Ebeling}, {Mullis}, \&
  {Tully}}]{Kocevski2007_CIZA}
{Kocevski}, D.~D., {Ebeling}, H., {Mullis}, C.~R., \& {Tully}, R.~B. 2007,
  \apj, 662, 224

\bibitem[{{Lane} {et~al.}(2014){Lane}, {Cotton}, {van Velzen}, {Clarke},
  {Kassim}, {Helmboldt}, {Lazio}, \& {Cohen}}]{Lane2014}
{Lane}, W.~M., {Cotton}, W.~D., {van Velzen}, S., {et~al.} 2014, \mnras, 440,
  327

\bibitem[{{Leahy} \& {Roger}(1998)}]{LeahyRoger1998}
{Leahy}, D.~A. \& {Roger}, R.~S. 1998, \apj, 505, 784

\bibitem[{{Lee} {et~al.}(2024){Lee}, {Pillepich}, {ZuHone}, {Nelson}, {Jee},
  {Nagai}, \& {Finner}}]{Lee2024}
{Lee}, W., {Pillepich}, A., {ZuHone}, J., {et~al.} 2024, \aap, 686, A55

\bibitem[{{Loi} {et~al.}(2017){Loi}, {Murgia}, {Govoni}, {Vacca}, {Feretti},
  {Giovannini}, {Carretti}, {Gastaldello}, {Girardi}, {Vazza}, {Concu},
  {Melis}, {Paladino}, {Poppi}, {Valente}, {Boschin}, {Clarke},
  {Colafrancesco}, {En{\ss}lin}, {Ferrari}, {de Gasperin}, {Gregorini},
  {Johnston-Hollitt}, {Junklewitz}, {Orr{\`u}}, {Parma}, {Perley}, \&
  {Taylor}}]{Loi2017}
{Loi}, F., {Murgia}, M., {Govoni}, F., {et~al.} 2017, \mnras, 472, 3605

\bibitem[{{Loi} {et~al.}(2020){Loi}, {Murgia}, {Vacca}, {Govoni}, {Melis},
  {Wittor}, {Kierdorf}, {Bonafede}, {Boschin}, {Brienza}, {Carretti}, {Concu},
  {Feretti}, {Gastaldello}, {Paladino}, {Rajpurohit}, {Serra}, \&
  {Vazza}}]{Loi2020}
{Loi}, F., {Murgia}, M., {Vacca}, V., {et~al.} 2020, \mnras, 498, 1628

\bibitem[{{Markevitch} {et~al.}(2005){Markevitch}, {Govoni}, {Brunetti}, \&
  {Jerius}}]{Markevitch2005_A520}
{Markevitch}, M., {Govoni}, F., {Brunetti}, G., \& {Jerius}, D. 2005, \apj,
  627, 733

\bibitem[{{Massey}(1951)}]{Massey1951}
{Massey}, F. J.~J. 1951, Journal of the American Statistical Association, 46,
  68

\bibitem[{{Mohan} \& {Rafferty}(2015)}]{Mohan2015}
{Mohan}, N. \& {Rafferty}, D. 2015, {PyBDSF: Python Blob Detection and Source
  Finder}, Astrophysics Source Code Library, record ascl:1502.007

\bibitem[{{Offringa} {et~al.}(2014){Offringa}, {McKinley}, {Hurley-Walker},
  {Briggs}, {Wayth}, {Kaplan}, {Bell}, {Feng}, {Neben}, {Hughes}, {Rhee},
  {Murphy}, {Bhat}, {Bernardi}, {Bowman}, {Cappallo}, {Corey}, {Deshpande},
  {Emrich}, {Ewall-Wice}, {Gaensler}, {Goeke}, {Greenhill}, {Hazelton},
  {Hindson}, {Johnston-Hollitt}, {Jacobs}, {Kasper}, {Kratzenberg}, {Lenc},
  {Lonsdale}, {Lynch}, {McWhirter}, {Mitchell}, {Morales}, {Morgan},
  {Kudryavtseva}, {Oberoi}, {Ord}, {Pindor}, {Procopio}, {Prabu}, {Riding},
  {Roshi}, {Shankar}, {Srivani}, {Subrahmanyan}, {Tingay}, {Waterson},
  {Webster}, {Whitney}, {Williams}, \& {Williams}}]{Offringa2014}
{Offringa}, A.~R., {McKinley}, B., {Hurley-Walker}, N., {et~al.} 2014, \mnras,
  444, 606

\bibitem[{{Ogrean} {et~al.}(2013){Ogrean}, {Br{\"u}ggen}, {R{\"o}ttgering},
  {Simionescu}, {Croston}, {van Weeren}, \& {Hoeft}}]{Ogrean2013XMM}
{Ogrean}, G.~A., {Br{\"u}ggen}, M., {R{\"o}ttgering}, H., {et~al.} 2013,
  \mnras, 429, 2617

\bibitem[{{Ogrean} {et~al.}(2014){Ogrean}, {Br{\"u}ggen}, {van Weeren},
  {R{\"o}ttgering}, {Simionescu}, {Hoeft}, \& {Croston}}]{Ogrean2014Chandra}
{Ogrean}, G.~A., {Br{\"u}ggen}, M., {van Weeren}, R., {et~al.} 2014, \mnras,
  440, 3416

\bibitem[{{Pasini} {et~al.}(2022){Pasini}, {Edler}, {Br{\"u}ggen}, {de
  Gasperin}, {Botteon}, {Rajpurohit}, {van Weeren}, {Gastaldello}, {Gaspari},
  {Brunetti}, {Cuciti}, {Nanci}, {di Gennaro}, {Rossetti}, {Dallacasa},
  {Hoang}, \& {Riseley}}]{Pasini2022}
{Pasini}, T., {Edler}, H.~W., {Br{\"u}ggen}, M., {et~al.} 2022, \aap, 663, A105

\bibitem[{{Pinzke} {et~al.}(2013){Pinzke}, {Oh}, \& {Pfrommer}}]{Pinzke2013}
{Pinzke}, A., {Oh}, S.~P., \& {Pfrommer}, C. 2013, \mnras, 435, 1061

\bibitem[{{Raja} {et~al.}(2024){Raja}, {Smirnov}, {Venturi}, {Rahaman}, \&
  {Yang}}]{Raja2024}
{Raja}, R., {Smirnov}, O.~M., {Venturi}, T., {Rahaman}, M., \& {Yang}, H. Y.~K.
  2024, \apj, 977, 83

\bibitem[{{Rajpurohit} {et~al.}(2022){Rajpurohit}, {van Weeren}, {Hoeft},
  {Vazza}, {Brienza}, {Forman}, {Wittor}, {Dom{\'\i}nguez-Fern{\'a}ndez},
  {Rajpurohit}, {Riseley}, {Botteon}, {Osinga}, {Brunetti}, {Bonnassieux},
  {Bonafede}, {Rajpurohit}, {Stuardi}, {Drabent}, {Br{\"u}ggen}, {Dallacasa},
  {Shimwell}, {R{\"o}ttgering}, {de Gasperin}, {Miley}, \&
  {Rossetti}}]{Rajpurohit2022_A2256}
{Rajpurohit}, K., {van Weeren}, R.~J., {Hoeft}, M., {et~al.} 2022, \apj, 927,
  80

\bibitem[{{Rajpurohit} {et~al.}(2020){Rajpurohit}, {Vazza}, {Hoeft}, {Loi},
  {Beck}, {Vacca}, {Kierdorf}, {van Weeren}, {Wittor}, {Govoni}, {Murgia},
  {Riseley}, {Locatelli}, {Drabent}, \&
  {Bonnassieux}}]{Rajpurohit2020_Toothbrush}
{Rajpurohit}, K., {Vazza}, F., {Hoeft}, M., {et~al.} 2020, \aap, 642, L13

\bibitem[{{Riseley} {et~al.}(2022){Riseley}, {Bonnassieux}, {Vernstrom},
  {Galvin}, {Chokshi}, {Botteon}, {Rajpurohit}, {Duchesne}, {Bonafede},
  {Rudnick}, {Hoeft}, {Quici}, {Eckert}, {Brienza}, {Tasse}, {Carretti},
  {Collier}, {Diego}, {Di Mascolo}, {Hopkins}, {Johnston-Hollitt}, {Keel},
  {Koribalski}, \& {Reiprich}}]{Riseley2022}
{Riseley}, C.~J., {Bonnassieux}, E., {Vernstrom}, T., {et~al.} 2022, \mnras,
  515, 1871

\bibitem[{{Schwarz}(1978)}]{Schwarz1978}
{Schwarz}, G. 1978, Annals of Statistics, 6, 461

\bibitem[{{Skillman} {et~al.}(2013){Skillman}, {Xu}, {Hallman}, {O'Shea},
  {Burns}, {Li}, {Collins}, \& {Norman}}]{Skillman2013}
{Skillman}, S.~W., {Xu}, H., {Hallman}, E.~J., {et~al.} 2013, \apj, 765, 21

\bibitem[{{Stroe} {et~al.}(2014{\natexlab{a}}){Stroe}, {Harwood}, {Hardcastle},
  \& {R{\"o}ttgering}}]{Stroe2014_Spectral_age_modelling}
{Stroe}, A., {Harwood}, J.~J., {Hardcastle}, M.~J., \& {R{\"o}ttgering}, H.
  J.~A. 2014{\natexlab{a}}, \mnras, 445, 1213

\bibitem[{{Stroe} {et~al.}(2014{\natexlab{b}}){Stroe}, {Rumsey}, {Harwood},
  {van Weeren}, {Rottgering}, {Saunders}, {Sobral}, {Perrott}, \&
  {Schammel}}]{Stroe2014_16GHz}
{Stroe}, A., {Rumsey}, C., {Harwood}, J.~J., {et~al.} 2014{\natexlab{b}},
  \mnras, 441, L41

\bibitem[{{Stroe} {et~al.}(2016){Stroe}, {Shimwell}, {Rumsey}, {van Weeren},
  {Kierdorf}, {Donnert}, {Jones}, {R{\"o}ttgering}, {Hoeft},
  {Rodr{\'\i}guez-Gonz{\'a}lvez}, {Harwood}, \& {Saunders}}]{Stroe2016}
{Stroe}, A., {Shimwell}, T., {Rumsey}, C., {et~al.} 2016, \mnras, 455, 2402

\bibitem[{{Stroe} {et~al.}(2013{\natexlab{a}}){Stroe}, {van Weeren}, {Intema},
  {R{\"o}ttgering}, {Br{\"u}ggen}, \& {Hoeft}}]{2013Stroe}
{Stroe}, A., {van Weeren}, R.~J., {Intema}, H.~T., {et~al.} 2013{\natexlab{a}},
  \aap, 555, A110

\bibitem[{{Stroe} {et~al.}(2013{\natexlab{b}}){Stroe}, {van Weeren}, {Intema},
  {R{\"o}ttgering}, {Br{\"u}ggen}, \& {Hoeft}}]{Stroe2013Discovery}
{Stroe}, A., {van Weeren}, R.~J., {Intema}, H.~T., {et~al.} 2013{\natexlab{b}},
  \aap, 555, A110

\bibitem[{{Van der Tol}(2009)}]{VanderTol2009PhDT}
{Van der Tol}, S. 2009, PhD thesis, Technical University of Delft, Netherlands

\bibitem[{{van Haarlem} {et~al.}(2013){van Haarlem}, {Wise}, {Gunst}, {Heald},
  {McKean}, {Hessels}, {de Bruyn}, {Nijboer}, {Swinbank}, {Fallows},
  {Brentjens}, {Nelles}, {Beck}, {Falcke}, {Fender}, {H{\"o}randel},
  {Koopmans}, {Mann}, {Miley}, {R{\"o}ttgering}, {Stappers}, {Wijers},
  {Zaroubi}, {van den Akker}, {Alexov}, {Anderson}, {Anderson}, {van Ardenne},
  {Arts}, {Asgekar}, {Avruch}, {Batejat}, {B{\"a}hren}, {Bell}, {Bell}, {van
  Bemmel}, {Bennema}, {Bentum}, {Bernardi}, {Best}, {B{\^\i}rzan}, {Bonafede},
  {Boonstra}, {Braun}, {Bregman}, {Breitling}, {van de Brink}, {Broderick},
  {Broekema}, {Brouw}, {Br{\"u}ggen}, {Butcher}, {van Cappellen}, {Ciardi},
  {Coenen}, {Conway}, {Coolen}, {Corstanje}, {Damstra}, {Davies}, {Deller},
  {Dettmar}, {van Diepen}, {Dijkstra}, {Donker}, {Doorduin}, {Dromer}, {Drost},
  {van Duin}, {Eisl{\"o}ffel}, {van Enst}, {Ferrari}, {Frieswijk}, {Gankema},
  {Garrett}, {de Gasperin}, {Gerbers}, {de Geus}, {Grie{\ss}meier}, {Grit},
  {Gruppen}, {Hamaker}, {Hassall}, {Hoeft}, {Holties}, {Horneffer}, {van der
  Horst}, {van Houwelingen}, {Huijgen}, {Iacobelli}, {Intema}, {Jackson},
  {Jelic}, {de Jong}, {Juette}, {Kant}, {Karastergiou}, {Koers}, {Kollen},
  {Kondratiev}, {Kooistra}, {Koopman}, {Koster}, {Kuniyoshi}, {Kramer},
  {Kuper}, {Lambropoulos}, {Law}, {van Leeuwen}, {Lemaitre}, {Loose}, {Maat},
  {Macario}, {Markoff}, {Masters}, {McFadden}, {McKay-Bukowski}, {Meijering},
  {Meulman}, {Mevius}, {Middelberg}, {Millenaar}, {Miller-Jones}, {Mohan},
  {Mol}, {Morawietz}, {Morganti}, {Mulcahy}, {Mulder}, {Munk}, {Nieuwenhuis},
  {van Nieuwpoort}, {Noordam}, {Norden}, {Noutsos}, {Offringa}, {Olofsson},
  {Omar}, {Orr{\'u}}, {Overeem}, {Paas}, {Pandey-Pommier}, {Pandey}, {Pizzo},
  {Polatidis}, {Rafferty}, {Rawlings}, {Reich}, {de Reijer}, {Reitsma},
  {Renting}, {Riemers}, {Rol}, {Romein}, {Roosjen}, {Ruiter}, {Scaife}, {van
  der Schaaf}, {Scheers}, {Schellart}, {Schoenmakers}, {Schoonderbeek},
  {Serylak}, {Shulevski}, {Sluman}, {Smirnov}, {Sobey}, {Spreeuw}, {Steinmetz},
  {Sterks}, {Stiepel}, {Stuurwold}, {Tagger}, {Tang}, {Tasse}, {Thomas},
  {Thoudam}, {Toribio}, {van der Tol}, {Usov}, {van Veelen}, {van der Veen},
  {ter Veen}, {Verbiest}, {Vermeulen}, {Vermaas}, {Vocks}, {Vogt}, {de Vos},
  {van der Wal}, {van Weeren}, {Weggemans}, {Weltevrede}, {White}, {Wijnholds},
  {Wilhelmsson}, {Wucknitz}, {Yatawatta}, {Zarka}, {Zensus}, \& {van
  Zwieten}}]{vanHaarlem2013}
{van Haarlem}, M.~P., {Wise}, M.~W., {Gunst}, A.~W., {et~al.} 2013, \aap, 556,
  A2

\bibitem[{{van Weeren} {et~al.}(2017){van Weeren}, {Andrade-Santos}, {Dawson},
  {Golovich}, {Lal}, {Kang}, {Ryu}, {Br{\`\i}ggen}, {Ogrean}, {Forman},
  {Jones}, {Placco}, {Santucci}, {Wittman}, {Jee}, {Kraft}, {Sobral}, {Stroe},
  \& {Fogarty}}]{vanWeeren2017NatAs}
{van Weeren}, R.~J., {Andrade-Santos}, F., {Dawson}, W.~A., {et~al.} 2017,
  Nature Astronomy, 1, 0005

\bibitem[{{van Weeren} {et~al.}(2019){van Weeren}, {de Gasperin}, {Akamatsu},
  {Br{\"u}ggen}, {Feretti}, {Kang}, {Stroe}, \& {Zandanel}}]{vanWeeren2019}
{van Weeren}, R.~J., {de Gasperin}, F., {Akamatsu}, H., {et~al.} 2019, \ssr,
  215, 16

\bibitem[{{van Weeren} {et~al.}(2011{\natexlab{a}}){van Weeren}, {Hoeft},
  {R{\"o}ttgering}, {Br{\"u}ggen}, {Intema}, \& {van
  Velzen}}]{vanWeeren2011_ZwCl0008}
{van Weeren}, R.~J., {Hoeft}, M., {R{\"o}ttgering}, H.~J.~A., {et~al.}
  2011{\natexlab{a}}, \aap, 528, A38

\bibitem[{{van Weeren} {et~al.}(2011{\natexlab{b}}){van Weeren}, {Intema},
  {R{\"o}ttgering}, {Br{\"u}ggen}, \& {Hoeft}}]{vanWeeren2011GMRT}
{van Weeren}, R.~J., {Intema}, H.~T., {R{\"o}ttgering}, H.~J.~A.,
  {Br{\"u}ggen}, M., \& {Hoeft}, M. 2011{\natexlab{b}}, \memsai, 82, 569

\bibitem[{{van Weeren} {et~al.}(2010){van Weeren}, {R{\"o}ttgering},
  {Br{\"u}ggen}, \& {Hoeft}}]{vanWeeren2010}
{van Weeren}, R.~J., {R{\"o}ttgering}, H. J.~A., {Br{\"u}ggen}, M., \& {Hoeft},
  M. 2010, Science, 330, 347

\bibitem[{{van Weeren} {et~al.}(2012){van Weeren}, {R{\"o}ttgering}, {Intema},
  {Rudnick}, {Br{\"u}ggen}, {Hoeft}, \& {Oonk}}]{vanWeeren2012_toothbrush}
{van Weeren}, R.~J., {R{\"o}ttgering}, H.~J.~A., {Intema}, H.~T., {et~al.}
  2012, \aap, 546, A124

\bibitem[{{van Weeren} {et~al.}(2021){van Weeren}, {Shimwell}, {Botteon},
  {Brunetti}, {Br{\"u}ggen}, {Boxelaar}, {Cassano}, {Di Gennaro},
  {Andrade-Santos}, {Bonnassieux}, {Bonafede}, {Cuciti}, {Dallacasa}, {de
  Gasperin}, {Gastaldello}, {Hardcastle}, {Hoeft}, {Kraft}, {Mandal},
  {Rossetti}, {R{\"o}ttgering}, {Tasse}, \& {Wilber}}]{vanWeeren2021extraction}
{van Weeren}, R.~J., {Shimwell}, T.~W., {Botteon}, A., {et~al.} 2021, \aap,
  651, A115

\bibitem[{{Vazza} \& {Botteon}(2024)}]{Vazza2024Galax}
{Vazza}, F. \& {Botteon}, A. 2024, Galaxies, 12, 19

\bibitem[{{Whittingham} {et~al.}(2024){Whittingham}, {Pfrommer}, {Werhahn},
  {Jlassi}, \& {Girichidis}}]{Whittingham2024}
{Whittingham}, J., {Pfrommer}, C., {Werhahn}, M., {Jlassi}, L., \&
  {Girichidis}, P. 2024, arXiv e-prints, arXiv:2411.11947

\bibitem[{{Wittor}(2023)}]{Wittor2023Univ}
{Wittor}, D. 2023, Universe, 9, 319

\bibitem[{{Wittor} {et~al.}(2021){Wittor}, {Ettori}, {Vazza}, {Rajpurohit},
  {Hoeft}, \& {Dom{\'\i}nguez-Fern{\'a}ndez}}]{Wittor2021}
{Wittor}, D., {Ettori}, S., {Vazza}, F., {et~al.} 2021, \mnras, 506, 396

\bibitem[{{Wittor} {et~al.}(2019){Wittor}, {Hoeft}, {Vazza}, {Br{\"u}ggen}, \&
  {Dom{\'\i}nguez-Fern{\'a}ndez}}]{Wittor2019}
{Wittor}, D., {Hoeft}, M., {Vazza}, F., {Br{\"u}ggen}, M., \&
  {Dom{\'\i}nguez-Fern{\'a}ndez}, P. 2019, \mnras, 490, 3987

\bibitem[{{Wittor} {et~al.}(2016){Wittor}, {Vazza}, \&
  {Br{\"u}ggen}}]{Wittor2016}
{Wittor}, D., {Vazza}, F., \& {Br{\"u}ggen}, M. 2016, Galaxies, 4, 71

\end{thebibliography}

\begin{appendix}
\section{Deconvolved profile shape comparison}
\label{app:profile_shape_analysis}

\begin{figure}[h!]
    \centering
    \includegraphics[width=0.95\columnwidth]{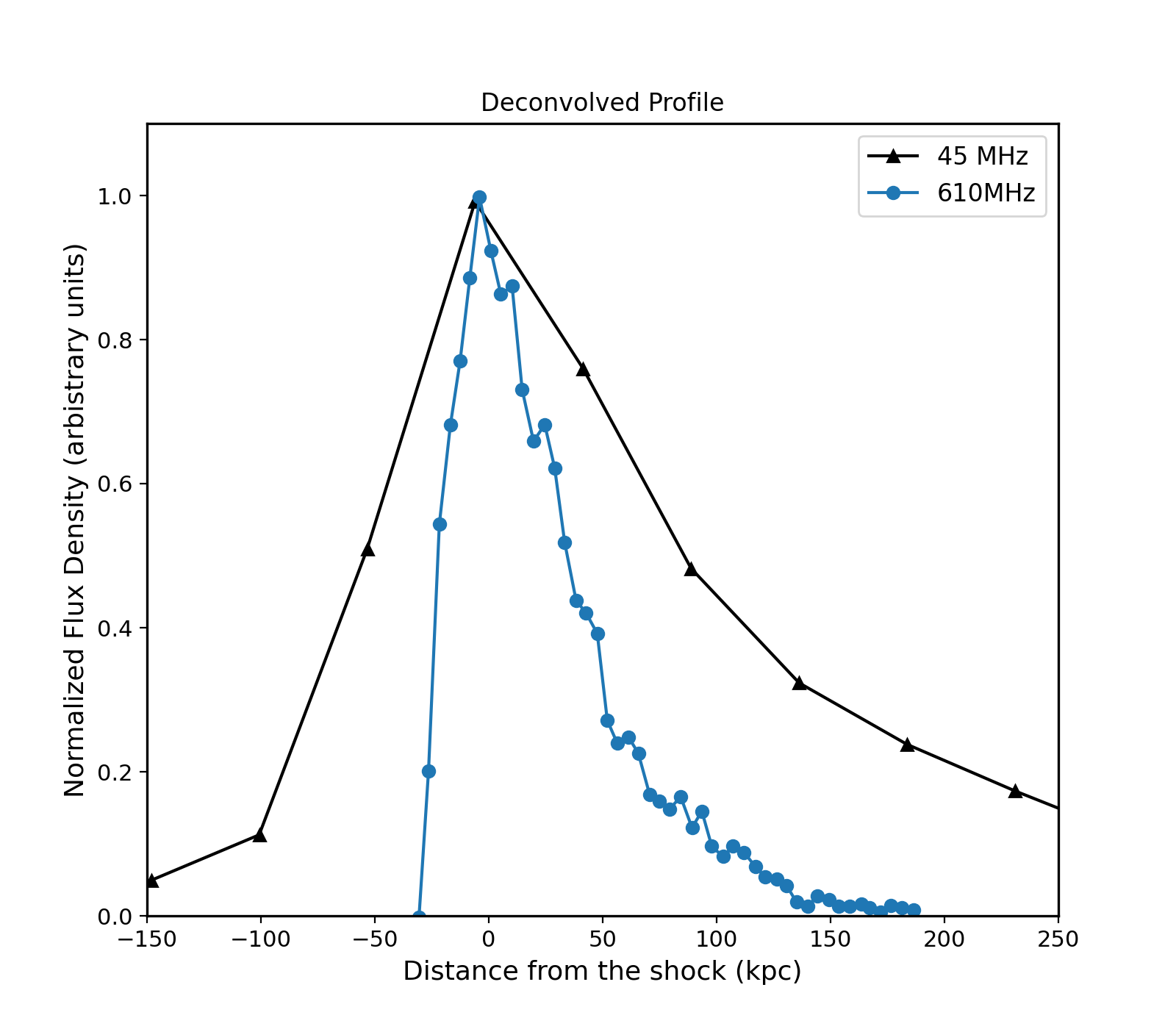}
    \caption{Comparison between full SB profile at 45 MHz (this work) and 610 MHz \citep{vanWeeren2010}. Data are shifted so that the peak of each curve corresponds to zero. We highlight that this is intended purely as a qualitative shape comparison.}
    \label{fig:deconv_prof_45_610}
\end{figure}

We compared the shape of the deconvolved profiles at 45 MHz (RN-full from \cref{sec: SB_profiles}) and 610 MHz \citep{vanWeeren2010} by computing some statistical metrics, such as the skewness and the 10-to-50 ratio.
For normally distributed data, the skewness should be about zero. A skewness value greater than zero means that there is more weight in the right tail of the distribution and vice versa.
This 10-to-50 ratio quantifies the relative extent of the wings compared to the core of the profile. It is calculated as the ratio of the profile width at 10\% of the maximum intensity to its full width at half maximum. A perfectly Gaussian profile has a value of 4.29. Ratios larger than 4.29 indicate broader wings, while smaller ratios suggest a steeper fall-off outside the core.
We also calculate the skewness test to determine if the skewness is close enough to zero. To do this, we test the null hypothesis that the skewness of the population that the sample was drawn from is the same as that of a corresponding normal distribution.
The null hypothesis is rejected when the p-value is low ($p < 0.05$), indicating that the distribution is significantly skewed and deviates from normality. Conversely, a high p-value would indicate a more symmetric form and that the skewness is not statistically different from that of a normal distribution.
The metrics results for the two curves shown in \cref{fig:deconv_prof_45_610} are reported in \cref{tab:skewness_comparison}.
\begin{table}[h!]
    \centering
    \begin{threeparttable}
        \caption{Comparison of RN deconvolved profiles shapes at 45 MHz and 610 MHz \citep{vanWeeren2010}.}
        \label{tab:skewness_comparison}
        \begin{tabular}{lcc}
            \toprule
            Metric & 45 MHz & 610 MHz \\
            \midrule
            Skewness & 0.63 & 0.91 \\
            p-value$^{*}$ & $3.25 \times 10^{-14}$ & 0.0097 \\
            10-to-50 Ratio & 2.73 & 2.18 \\
            \bottomrule
        \end{tabular}
        \begin{tablenotes}
            \footnotesize
            \item $^{*}$The p-value for the skewness-hypothesis test.
        \end{tablenotes}
    \end{threeparttable}
\end{table}

Both profiles show a moderate (between 0.5 and 1) positive skewness. The 45 MHz profile is more symmetric, with a skewness between -0.5 and 0.5.
Despite this, according to the skewness test, both profiles show significant evidence of deviations from Gaussianity (p-value $\ll 0.05$).
We conclude that the 45 MHz profile is broader (larger than 10-to-50 ratio) and more symmetric (lower skewness) than the 610 MHz profile. On the other hand, the 610 MHz profile is more peaked, has steeper wings, and is more asymmetric.

\section{Profile modelling}
\label{app:profile_model_analysis_psi18}

\begin{figure}[h!]
    \centering
    \includegraphics[width=0.95\columnwidth]{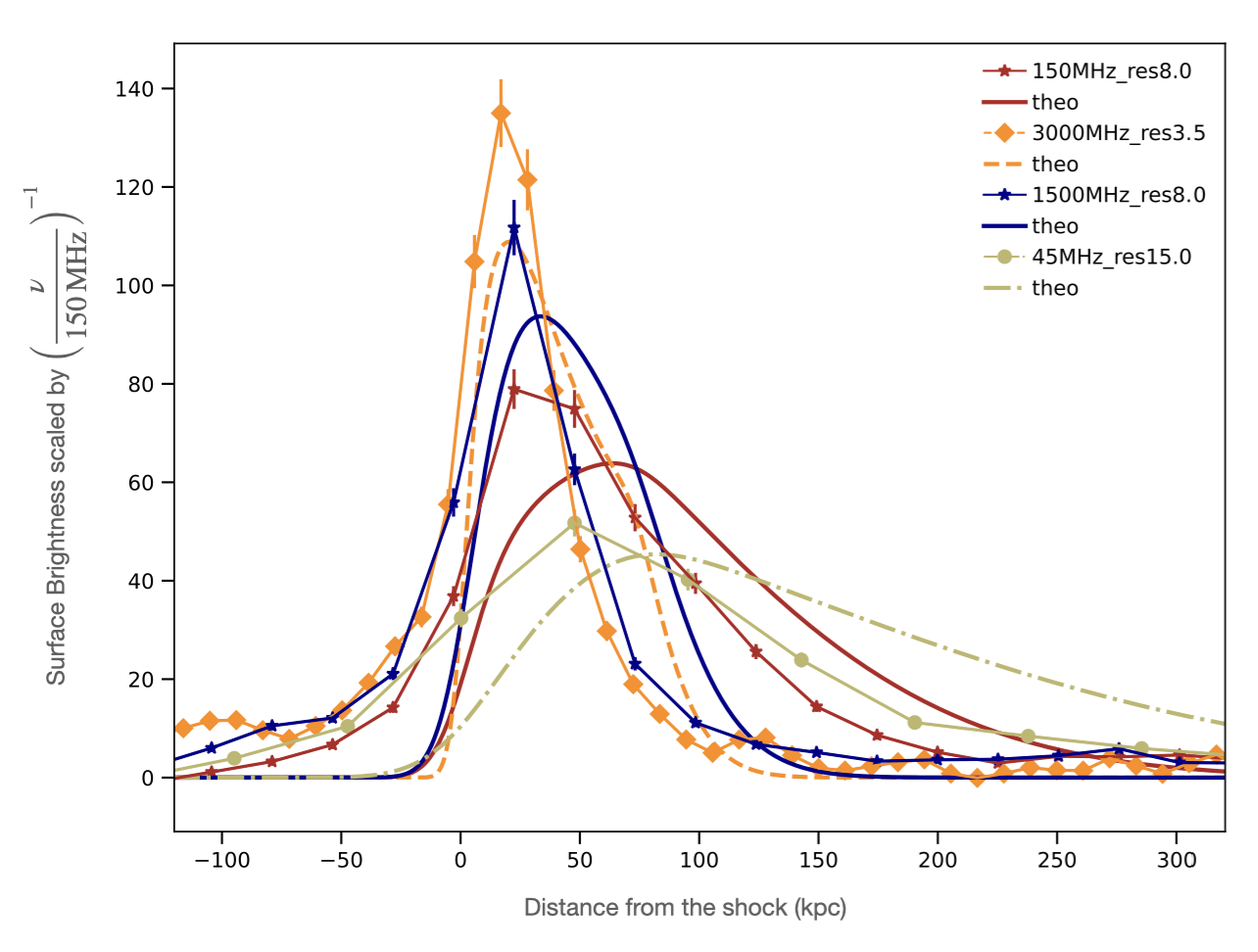}
    \caption{Surface brightness profile across the eastern half of RN. Same as \cref{fig:SB_profile_model} but with constant $\rm B=3\,\mu G$ and $\rm \Psi = 18^{\circ}$ projection.}
    \label{fig:SB_profile_model_18deg}
\end{figure}

In this section, we show the effect of a bigger opening angle $\rm \Psi$. Even just a slight change (from $12^{\circ}$ to $18^{\circ}$) drastically changes the downstream profile, making the model unfeasible (see \cref{fig:SB_profile_model_18deg}).
The effect is particularly visible in the high-frequency high-resolution profiles (1.5 and 3 GHz, blue and orange respectively).

\end{appendix}

\end{document}